\documentclass[
  aps,
  prd,
  nofootinbib,
  onecolumn,
  twocolumn,
  preprintnumbers,
  superscriptaddress,
  groupedaddress,
  floatfix,
]{revtex4-2}

\usepackage{silence} 
\usepackage[
  pdftex,
  colorlinks=true,
  pdfpagemode=UseNone,
  bookmarksopen=true,
]{hyperref}

\usepackage[utf8]{inputenc}
\usepackage{amsmath} 
\usepackage{amssymb} 
\usepackage{bm} 
\usepackage{cancel} 
\usepackage{calc}
\usepackage{color}
\usepackage{dcolumn} 
\usepackage{graphicx}
\usepackage[normalem]{ulem} 

\newcommand{\LLNL}{Nuclear and Chemical Sciences Division, Lawrence Livermore National Laboratory, Livermore, CA 94550, USA}

\newif\ifdocomment
\docommentfalse
\docommenttrue
\ifdocomment
\newcommand{\comment}[1]{\textcolor{red}{[#1]}}
\newcommand{\highlight}[1]{\textcolor{red}{#1}}
\newcommand{\str}[1]{}
\newcommand{\postsubmissionStrike}[1]{\sout{#1}}

\else
\newcommand{\comment}[1]{\PackageError{comment}{unresolved comment}{}}
\newcommand{\highlight}[1]{\PackageError{comment}{unresolved highlight}{}}
\newcommand{\str}[1]{}
\newcommand{\postsubmissionStrike}[1]{}

\fi

\newcommand{\kmax}{\ensuremath{k_{\rm max}}}
\newcommand{\tc}{\ensuremath{t_{c}}}
\newcommand{\tz}{\ensuremath{t_{0}}}

\newcommand{\z}{\ensuremath{z}}
\newcommand{\zexp}{\z\text{ expansion}}
\newcommand{\p}{\ensuremath{p}}
\newcommand{\pvalue}{\p\text{ value}}
\newcommand{\Qtcom}{\ensuremath{Q^{2}_{\rm compare}}}
\newcommand{\Qtmin}{\ensuremath{Q^{2}_{\rm min}}}
\newcommand{\Qtanc}{\ensuremath{Q^{2}_{\rm anchor}}}
\newcommand{\GeV}{\text{GeV}}
\newcommand{\GeVsq}{\ensuremath{\GeV^{2}}}
\newcommand{\fm}{\text{fm}}
\newcommand{\fmsq}{\ensuremath{\fm^{2}}}

\newcommand{\rasq}{\ensuremath{r_{A}^{2}}}

\newcommand{\cov}{\text{cov}}
\newcommand{\naive}{\text{naive}}
\newcommand{\derated}{\text{derated}}

\newcommand{\rfr}[1]{Ref.~\cite{#1}}
\newcommand{\eqn}[1]{Eq.~(\ref{#1})}
\newcommand{\fgr}[1]{Fig.~\ref{#1}}
\newcommand{\tbl}[1]{Table~\ref{#1}}
\newcommand{\sct}[1]{Sec.~\ref{#1}}
\newcommand{\rfrs}{Refs.}
\newcommand{\eqns}{Eqs.}

\newcommand{\scts}{Secs.}

\newcommand{\minerva}{\text{MINER}\ensuremath{\nu}\text{A}}

\newcommand{\rqcd}{\text{RQCD}}
\newcommand{\nme}{\text{NME}}
\newcommand{\pndme}{\text{PNDME}}
\newcommand{\mainz}{\text{Djukanovic {\it et al.}}}
\newcommand{\etm}{\text{ETM}}

\newcommand{\phz}{\phantom{\ensuremath{0}}}
\newcommand{\phm}{\phantom{\ensuremath{-}}}
\newcommand{\pha}{\phantom{\ensuremath{*}}}
\newcommand{\nustattools}{\text{\tt nustattools}}
\newcommand{\stat}{\text{stat}}
\newcommand{\syst}{\text{syst}}
\newcommand{\total}{\text{total}}

\begin{document}
\preprint{LLNL-JRNL-2014602}
\title{The Nucleon Axial Form Factor from Averaging Lattice QCD Results}

\author{Aaron~S.~Meyer}
\email{asmeyer.physics@gmail.com}
\email{meyer54@llnl.gov}
\affiliation{\LLNL}

\date{\today}

\begin{abstract}
Flagship neutrino oscillation experiments depend on
 precise and accurate theoretical knowledge of
 neutrino-nucleon cross sections across a variety of energies
 and interaction mechanisms.
Key ingredients to the amplitudes that make up these cross sections
 are parameterized form factors.
The axial form factor describing a weak interaction with a nucleon
 is part of one of the primary neutrino-nucleon interaction mechanisms,
 quasielastic scattering, yet this form factor is uncertain and
 its precision is limited by the availability of data for a
 neutrino scattering with nucleons or small nuclear targets.
Lattice Quantum Chromodynamics (LQCD) now offers another approach
 for obtaining mathematically rigorous constraints of the axial form factor
 from theoretical calculations with complete systematic error budgets.
In this work, strategies for averaging LQCD results are explored,
 including both a random sampling of form factor values across momentum transfers
 as well as an averaging strategy based on analytic calculations of form factor derivatives.
Fits to \zexp{} parameterizations are reported and compared
 against neutrino-hydrogen and neutrino-deuterium scattering data.
\end{abstract}
\maketitle

\section{Introduction}

Experiments that measure neutrino scattering and interactions
 are evolving from an era of discovery to one of precision.
Flagship neutrino oscillation experiments~\cite{
 JUNO:2015sjr,
 DUNE:2015lol,
 Hyper-Kamiokande:2018ofw,
 NOvA:2007rmc,
 DUNE:2020ypp} are preparing for their
 initial data taking phases, where they will detect an unprecedented
 number of neutrino interactions.
These measurements will help to close the gaps in scientific understanding
 of the parameters that govern neutrino flavor oscillation.

To help achieve and surpass the goals put forth by neutrino experiments,
 sufficient theoretical support must also be provided that can
 aid in interpretation of detected events.
Neutrino beams contain a broad spectrum of neutrino energies,
 preventing accurate quantification of the neutrino energy on an event-by-event basis.
Neutrino interactions must instead be described as statistical distributions
 and the corresponding neutrino energies reconstructed from those distributions.
This process requires precise and accurate knowledge of neutrino cross sections
 across a range of neutrino energies and interaction topologies.

The cross sections describing neutrino interactions with nuclear matter
 are constructed from weak interaction amplitudes.
Theoretical characterization of these amplitudes depend largely on
 free nucleon matrix elements, which have historically been obtained
 from scattering with small nuclear targets.
Parameterized by form factors that depend upon
 the four-momentum transfer squared ($Q^{2}$),
 the vector contributions to low-energy neutrino
 scattering with a nucleon target can be precisely obtained from
 electron-proton scattering data.
The axial contributions, on the other hand, can only be obtained from
 measurements sensitive to weak matrix elements
 and therefore relatively low-statistics experiments.
Measurements from bubble chambers containing deuterium fills~\cite{
 Mann:1973pr,
 Barish:1977qk,
 Barish:1978pj,
 Miller:1982qi,
 Baker:1981su,
 Kitagaki:1983px,
 Wachsmuth:1979th,
 Barlag:1984uga,
 Allasia:1990uy}
 have been the most constraining on the primary source
 of uncertainty for neutrino cross section, the axial form factor.
Other works using pion electroproduction~\cite{
 Hilt:2013fda,
 A1:1999kwj,
 Blomqvist:1996tx}
 and muon capture~\cite{
 Czarnecki:2007th,
 MuCap:2012lei,
 MuCap:2015boo,
 Hill:2017wgb}
 are also sensitive to the axial form factor.
More recently, a result for the axial form factor obtained
 from antineutrino-hydrogen scattering has also been
 published~\cite{MINERvA:2022vmb,MINERvA:2023avz}.

\newcommand{\lqcd}{\text{LQCD}}
Due to advances in computational techniques
 and the growth of computing power with Moore's Law,
 theoretical computations of weak matrix elements with neutrinos
 are now possible using Lattice Quantum Chromodynamics (\lqcd).
\lqcd{} calculations access the properties of hadrons by means
 of numerical evaluations of the path integral formulation,
 providing mathematically rigorous predictions and systematically
 improvable uncertainties on hadronic quantities such as masses
 and matrix elements.
Though experiments face difficulties measuring neutrino scattering
 with free nucleon targets, calculating the axial matrix elements
 with \lqcd{} is straightforward.
This makes the use of \lqcd{} an appealing candidate as a source
 of constraint for neutrino cross sections.

Historically, \lqcd{} calculations have had difficulty reproducing
 the axial coupling, $g_{A}$.
This value is precisely known from neutron beta decay measurements~\cite{PDG2024}
 and therefore a benchmark quantity for \lqcd{} calculations with nucleon states.
Past \lqcd{} calculations underpredicted $g_{A}$ by as much as 20\%~\cite{Kronfeld:2019nfb}.
This was later attributed to poorly-constrained pion-nucleon excited state contributions
 that were enhanced by the axial current
 \cite{
 Tiburzi:2015tta,
 Tiburzi:2015sra,
 Bar:2017kxh,
 Bar:2017gqh,
 Bar:2018akl,
 Bar:2018xyi,
 Bali:2018qus,
 Jang:2019vkm,
 Bar:2019gfx,
 Bar:2019zkx,
 RQCD:2019jai}.
With a better understanding of the relevant contamination,
 \lqcd{} calculations of the axial coupling are now able to reproduce
 the experimental value to within 1\%~\cite{
 Chang:2018uxx,
 Bali:2023sdi,
 Gupta:2024qip,
 Djukanovic:2024krw,
 Alexandrou:2024ozj,
 Hall:2025ytt}.

With relevant expertise and demonstrable control of the axial coupling,
 \lqcd{} collaborations have started reporting calculations of the
 axial form factor $Q^{2}$ dependence.
Several collaborations have released results within the past half decade
 using a nontrivial variety of quark actions and analysis techniques.
These calculations all pass internal consistency checks of the axial form factor
 against the pseudoscalar and induced pseudoscalar form factors
 by making use of the generalized Goldberger-Treimann relation
 and pion pole dominance ansatz~\cite{
 Gupta:2017dwj,
 Bali:2018qus,
 Jang:2019vkm,
 RQCD:2019jai,
 Ruso:2022qes}.
The collaborations also agree well with each other,
 providing a consistent narrative from the \lqcd{} community.

Surprisingly, the axial form factor predictions obtained from
 \lqcd{} calculations exhibit a significant tension at high $Q^{2}$
 compared to the axial form factor constraints originating from
 deuterium bubble chamber experiments~\cite{
 Meyer:2016oeg,
 Meyer:2022mix,
 Tomalak:2023pdi}
These changes suggest that the quasielastic cross section
 should be enhanced by as much as 30\% over the previous
 deuterium prediction, well outside of the 10\% uncertainty.
Within the same timeframe that the \lqcd{} results appeared,
 accompanying evidence corroborating the need for an enhanced
 quasielastic cross section came in the form of Monte Carlo tunes
 to neutrino-nucleus data~\cite{MicroBooNE:2021ccs,GENIE:2022qrc}
 as well as a recent measurement of antineutrino-proton scattering~\cite{
 MINERvA:2022vmb,
 MINERvA:2023avz}.

With several independent \lqcd{} results already available,
 one might wonder how much the constraints on the axial form factor
 can be improved by averaging these results.
This work provides such a prediction of the axial form factor
 by combining calculations from five different collaborations
 over four unique sets of gauge ensembles.
These calculations provide nontrivial checks of consistency among
 each other and should be viewed as independent results.
Attention is paid to the correlations between results that make
 use of the same gauge ensembles.
This work was prepared in conjunction with a sister paper~\cite{Meyer:inprep},
 where many details about the experimental connections can be found.

This document is organized as follows.
\sct{sec:fit_details} discusses details associated with parameterizing the form factors,
 the construction of a sufficient set of fit residuals, and how to handle correlations
 between \lqcd{} results.
\sct{sec:lqcd_result_summary} gives the details of the \lqcd{} results that
 were used in this work and lists the parameters of the form factors
 that go into the average.
\sct{sec:fit_results} gives details about the fits, tests of systematics effects,
 and consistency checks between different fit methods.
Several parameterizations of the form factor under different assumptions are reported.
Some concluding remarks are given in \sct{sec:discussion}.

\section{Fit Details}
\label{sec:fit_details}

This section describes the details of how the fits to \lqcd{} results
 are carried out.
\sct{sec:parameterizations} enumerates the relevant parameterizations
 that appear in the literature and in this work.
\sct{sec:fit_methods} gives details on how fits are constructed,
 including the residuals that are optimized over
 and the assumed covariances on these residuals.
\sct{sec:unknown_covariances} describes strategies for addressing
 unknown correlations in the \lqcd{} parameterizations due to shared
 gauge ensembles.

\subsection{Form Factor Parameterizations}
\label{sec:parameterizations}

Historically, several parameterizations have been used
 to describe the shape and magnitude of the form factor.
Although this work primarily focuses on the \zexp{} parameterization
 of \sct{sec:parameterization_zexp},
 the dipole parameterization in \sct{sec:parameterization_dipole}
 is also explored due to its widespread use.

\subsubsection{Dipole Parameterization}
\label{sec:parameterization_dipole}

The historical choice of parameterization used for fitting
 the axial form factor has been the dipole
 parameterization~\cite{LlewellynSmith:1971uhs},
 given by
\begin{align}
 F_{A}(Q^{2}) = g_{A} \big[ 1 + Q^{2}/M_{A}^{2} \big]^{-2}.
\end{align}
This parameterization has two free parameters:
 $g_{A}$, the axial coupling, which is fixed by precise neutron beta decay experiments;
 and $M_{A}$, also known as the axial mass parameter.
The dipole form factor is also the Fourier transform
 of a simple exponentially decaying function in three dimensions.

One of the reasons this form factor has persisted
 for so long is that this parameterization
 has an asymptotic falloff proportional
 to $Q^{-4}$ as $Q^{2}\to\infty$,
 in agreement with the expectations from perturbative QCD~\cite{Lepage:1980fj}.
However, this limiting behavior only becomes relevant for $Q^{2}$ well
 outside of the range probed by neutrino experiments.
Even if the form factor has the correct power-law falloff in the large $Q^{2}$ limit,
 there is no guarantee that the prefactor multiplying
 that asymptotic falloff is correct.

The dipole parameterization of the axial form factor
 is overconstrained by neutrino scattering data.
Given the precise constraints on $g_{A}$,
 the axial form factor has effectively one free parameter
 to constrain with neutrino scattering data.
Different $Q^{2}$ ranges result in different best-fit values for $M_{A}$,
 and fits to the full range typically result in poor fit quality
 and underestimated uncertainties.
This resulted in an ``axial mass problem,''
 where values of $M_{A}$ reported by different sources disagreed
 by more than their quoted uncertainties.

To capture a more realistic range of uncertainties for the form factor,
 $M_{A}$ has historically been considered with an inflated 10\% relative uncertainty.
This choice still assumes 100\% correlation between different
 values of $Q^{2}$, providing no flexibility in the overall shape.
More well-motivated choices with alternative parameterizations,
 for example the \zexp, result in more realistic uncertainties
 that are an order of magnitude larger than those of the dipole prediction.
At least part of this larger uncertainty comes from allowing more flexibility
 in the form factor shape across $Q^{2}$.

\subsubsection{\texorpdfstring{$z$}{z} Expansion Parameterization}
\label{sec:parameterization_zexp}

The \zexp{} parameterization~\cite{Hill:2010yb,Bhattacharya:2011ah}
 is a conformal mapping that takes $Q^{2}$ to a small expansion parameter $z$
 for the entire kinematic range probed by quasielastic scattering.
The definition of the conformal mapping is
\begin{align}
 z = \frac{\sqrt{\tc+Q^{2}} - \sqrt{\tc-\tz}}{\sqrt{\tc+Q^{2}} + \sqrt{\tc-\tz}}.
\end{align}
With this conformal mapping, the form factor can be written as a power series in $z$,
\begin{align}
 F_{A}(z) = \sum_{k=0}^{\infty} a_{k} z^{k}.
 \label{eq:z_expansion}
\end{align}
In practice, the sum is taken up to some finite order \kmax{}
 that is determined by the fit range and fidelity of the data.
The parameter \tc{} is bounded by the particle production threshold
 for the axial current interaction, at $\tc\leq(3M_{\pi})^{2}$.
The last remaining parameter, \tz, fixes the value of $Q^{2}$
 for which $z(Q^{2}=-\tz) = 0$ is satisfied.
The value of \tz{} can be chosen for convenience
 and typically is chosen to minimize the maximum
 value of $z$ over the entire kinematic range of interest.

The shape of the form factor can additionally be constrained
 with sum rules that enforce desirable behavior in the
 asymptotic $Q^{2}\to\infty$ limit.
These sum rules are obtained by solving for derivatives of the form factor
 with respect to $z$ taken as $z\to1$ ($Q^{2}\to\infty$),
\begin{align}
 \sum_{k=n}^{\kmax} k (k-1) \dots (k-n+1) a_{k} = 0 , n \in \{0,1,2,3\}.
 \label{eq:sum_rules}
\end{align}
These sum rules enforce the conditions
\begin{align}
 \lim_{Q^{2}\to\infty} \biggr[ \frac{\partial}{\partial Q^{2}} \biggr]^{n} F_{A}(Q^{2}) = 0.
\end{align}
Given the coefficients $a_{k}$ up to order $\kmax-4$,
 the sum rules provide an exact relation to solve for
 the last 4 remaining coefficients.
Imposing sum rules as a constraint during fits to the \zexp{}
 does not significantly affect the fit results,
 typically resulting in a smaller uncertainty
 for the high-$Q^{2}$ range beyond the data and central values
 shifted within the $1\sigma$ uncertainty band.

The constraints on $g_{A}$ from neutrino scattering and
 from \lqcd{} are not competitive with constraints
 from neutron beta decay measurements.
Instead of treating the value of $g_{A}$ as a fit parameter,
 an additional constraint is imposed to fix the intercept
\begin{align}
 F_{A}(Q^{2}=0) = g^{\beta}_{A}.
 \label{eq:ga_constraint}
\end{align}
This constraint is enforced by fixing $a_{0}$
 to reproduce $g_{A}^{\beta}=1.2754$, taken from PDG~\cite{PDG2024}.
A summary of the full set of parameters
 adopted in this work is given in \tbl{tab:fitparameters}.

\begin{table}[btp]
\begin{tabular}{c|r}
Parameter & Value \\
\hline
$g_{A}^{\beta}$~\cite{PDG2024} & $1.2754$ \\
\tc & $9\cdot(0.134~\GeV)^{2}$ \\
\tz & $-0.50~\GeVsq$ \\
\# sum rules & 4
\end{tabular}
\caption{
A list of fixed parameters needed to fit
 the axial form factor in this work.
The first column gives the fit parameter,
 and the second column gives the value that
 was used for that parameter in this work.
\label{tab:fitparameters}
}
\end{table}

\subsection{Fit Methods}
\label{sec:fit_methods}

A set of residuals must be defined in order to construct
 a $\chi^{2}$ loss function to optimize.
There is more than one way to build a set of residuals
 and so various choices must be considered in order to ensure
 that the fits are not sensitive to the particular choices that were used.
Once a meaningful set of residuals $r_{i}$ has been decided upon,
 the loss function
\begin{align}
 \chi^{2} = \sum_{ij} r_{i}(\bm{a}) \Big[ C^{\bm{r}}_{ij} \Big]^{-1} r_{j}(\bm{a})
\end{align}
 can be minimized to find a set of best-fit parameters.
In this case, the parameters $\bm{a}$ to minimize are exactly
 the \zexp{} coefficients $a_{k}$ in \eqn{eq:z_expansion}.
For constructing a covariance of the residuals,
 it is simple to use Gaussian error propagation to define a residual covariance
\begin{align}
 C^{\bm{r}}_{ij}
 =
 \sum_{k\ell}
 \big[ \nabla_{\alpha_{k}} r_{i} \big]
 C^{\bm{\alpha}}_{k\ell}
 \big[ \nabla_{\alpha_{\ell}} r_{j} \big]
 \label{eq:residual_covariances}
\end{align}
 using the covariance $C^{\bm{\alpha}}$ over the
 input parameters $\alpha_{k}$.

Two fit methods, leading to different residuals,
 will be employed in this work.
\sct{sec:fitmethod_derivatives} describes a fit using the central values and derivatives
 of the form factor at specific values of $Q^{2}$.
As an alternative, \sct{sec:fitmethod_sampled}
 considers fits where the full $Q^{2}$ range of the form factor
 is stochastically sampled and fit.

\subsubsection{Fit to Derivatives}
\label{sec:fitmethod_derivatives}

The axial form factor parameterizations from \lqcd{} results
 can be combined analytically by fitting to the form factor central
 value and its derivatives at fixed values of $Q^{2}$,
\begin{align}
 \biggr[ \frac{\partial}{\partial Q^{2}} \biggr]^{n} F_{A}(Q^{2}) , \; n \geq 0 .
\end{align}
With this choice, the set of residuals then becomes
\begin{align}
 r_{i} =
 \biggr[ \frac{\partial}{\partial Q^{2}} \biggr]^{n_i}
 \biggr[ F_{A}(Q^{2};\bm{\alpha}_{(i)}) - F_{A}(Q^{2};\bm{a}) \biggr]
 \biggr|_{Q^{2}=Q^{2}_{i}}
 \label{eq:residuals_derivatives}
\end{align}
 for choices of derivative order $n_{i}$,
 $Q^{2}$ evaluation point $Q^{2}_{i}$,
 and an input parameterization with coefficients $\bm{\alpha}_{(i)}$.
The evaluation point $Q^{2}=-\tz$ is particularly convenient for the
 \zexp{} parameterization because this choice yields the relations
\begin{align}
 F_{A}(Q^{2}=-\tz) &= a_{0} , \nonumber\\
 \frac{\partial}{\partial Q^{2}} F_{A} \biggr|_{Q^{2}=-\tz} &=
 a_{1} \frac{\partial z}{\partial Q^{2}} .
 \label{eq:fit_matching_condition}
\end{align}
All terms proportional to a power of $z$ vanish due to the requirement that
 $z(Q^{2}=-\tz) = 0$.
Higher-order derivatives of the form factor yield linear combinations
 of the $a_{k}$ multiplying derivatives of $z$,
 such as
\begin{align}
 \biggr[ \frac{\partial}{\partial Q^{2}} \biggr]^{2} F_{A} \biggr|_{Q^{2}=-\tz} &=
 2 a_{2} \biggr[ \frac{\partial z}{\partial Q^{2}} \biggr]^{2}
 + a_{1} \biggr[ \frac{\partial}{\partial Q^{2}} \biggr]^{2} z .
\end{align}
Although the nominal choice will be to compute the form factor and its
 derivatives at $Q^{2}=-\tz$, other choices will be explored in this work.

There are several advantages to fitting derivatives of the form factor.
Although this fit procedure is formulated in a way that is convenient
 for comparing \zexp{} parameterizations,
 the dependence only on form factor derivatives with respect to $Q^{2}$
 means the procedure can be applied to any parameterization.
The \zexp{} also exhibits linear behavior over a large range of parameter values,
 meaning that most of the form factor shape is captured in these first two derivatives
 even when they are computed only at a single value of $Q^{2}$.
Since most \lqcd{} parameterizations do not constrain more than 3 parameters,
 the assumption is that the last parameter is absorbing contributions from
 other higher-order terms as well, and therefore derivatives beyond
 first order are not expected to be well constrained.
These effects are expected to be primarily absorbed by the coefficients
 relegated to the sum rules, with only small effects seen
 in the lower order fit coefficients.
Another advantage of this formulation is that the form factor derivatives
 and uncertainty propagation can be computed analytically provided
 that the number of residuals for any \zexp{} parameterization does not exceed
 the number of \zexp{} coefficients.
The bulk of the form factor shape can therefore be extracted
 from just a few residuals with this well-defined fit procedure.

There are also disadvantages to fitting form factor derivatives
 as recommended in this section.
There is ambiguity in how to select the best momentum transfer value
 (or values) at which to evaluate the form factors,
 leading to a potential for bias.
For this, there at least exists the natural expansion point
 at $Q^{2}=-\tz$ that leads to \eqn{eq:fit_matching_condition},
 providing some measure of guidance.
As another drawback, care must be taken to avoid constructing
 sets of residuals that are not linearly independent,
 which can lead to an uninvertible covariance matrix.
This can be especially troublesome in nontrivial circumstances,
 for instance if all of the results are evaluated at the same $Q^{2}$,
 or if derivatives are always evaluated at $Q^{2}=-\tz$
 and the order of the expansion is greater than the number of included derivatives.
In such circumstances, one can run into uninvertible covariance matrices
 even for \kmax{} less than the number of degrees of freedom.
These deficiencies will be explored in the tests in \sct{sec:fits_derivatives}.

In general, $g_{A}$ is an output of \lqcd{} calculations rather than an input.
For this reason, the constraint on $g_{A}$ in \eqn{eq:ga_constraint}
 is not typically employed in the reported \lqcd{} results,
 instead only taking their own constraint on $g_{A}$ to fix the form factor intercept.
To match onto the results from neutrino scattering in a sister paper~\cite{Meyer:inprep},
 the constraint of \eqn{eq:ga_constraint} will be used when fitting \lqcd{} results here.
This introduces a complication.
Some \lqcd{} results use $\tz=0$ for simplicity,
 meaning that the matching condition in \eqn{eq:fit_matching_condition}
 should be performed at $Q^{2}$ where the form factor is not allowed to vary.
As a consequence, the residual for the form factor central value computed
 at $Q^{2}=0$ will be independent of all fit parameters.
Since higher-order $Q^{2}$ derivatives are independent of
 the \zexp{} coefficient $a_{0}$, matching the fits at $Q^{2}=0$
 then potentially drops the most strongly constrained \zexp{} coefficient
 from these parameterizations.
The resolution is simple: instead of computing the form factor
 and its derivatives at $Q^{2}=0$, the form factor matching
 is instead performed at some finite nonzero $Q^{2}$.
This evaluation point will be varied and selected
 to maximize the uncertainty across $Q^{2}$.

\subsubsection{Fit to Sampled Form Factor}
\label{sec:fitmethod_sampled}

A more straightforward way to fit the form factor is to sample
 the form factor central values over a fixed set of $Q^{2}$ points.
The residuals are then simply
\begin{align}
 r_{i} =
 \biggr[ F_{A}(Q^{2};\bm{\alpha}_{(i)}) - F_{A}(Q^{2};\bm{a}) \biggr]
 \biggr|_{Q^{2}=Q^{2}_{i}} ,
 \label{eq:residuals_sampled}
\end{align}
 defined with sampled $Q^{2}$ values $Q^{2}_{i}$
 and input parameters $\alpha_{i}$
 like in \eqn{eq:residuals_derivatives}.
The typical strategy is to choose a set of fixed, evenly-spaced set of $Q^{2}$
 points and use the same $Q^{2}$ values for all parameterizations
 that are being averaged.

Rather than rely on Gaussian uncertainty propagation,
 random values of the form factor parameters are selected respecting
 the covariances between the parameters.
The covariances are used to generate a set of stochastically
 sampled form factor curves over the full range of $Q^{2}$
 using each parameterization.
For each of these samples, the form factor is fit
 and the stochastic samples of the results are used to generate
 a covariance on the final fit parameters.
The residuals in \eqn{eq:residuals_sampled}
 are assumed to be uncorrelated to avoid the problem
 of an uninvertible covariance matrix, since the number
 of $Q^{2}$ values exceeds the number of input \lqcd{} parameters.

\subsubsection{Translated \zexp{}}
\label{sec:zexp_translated}

The observation of \eqn{eq:fit_matching_condition} hints at a scheme
 for comparing \zexp{} parameterizations that have been generated
 with different choices of \tc, \tz, and \kmax.
Given a set of \zexp{} coefficients for one parameterization,
 another parameterization can be obtained by computing derivatives
 with respect to $z$,
\begin{align}
 a_{k} = \frac{1}{k!} \frac{d^{k}F_{A}(z)}{dz^{k}} \Biggr|_{z=0} .
\end{align}
To compute these derivatives, chain rule is applied to convert
 $z$ derivatives to derivatives with respect to $Q^{2}$,
\begin{align}
 \frac{d}{dz} \to \frac{dQ^{2}}{dz} \frac{d}{dQ^{2}} .
\end{align}
Both sets of derivatives,
 namely the derivatives of $Q^{2}$ with respect to $z$
 and the derivatives of the form factor with respect to $Q^{2}$,
 can be computed analytically for \zexp{} parameterizations.
The form factor derivatives with respect to $Q^{2}$
 are computed at the point $Q^{2}=-\tz$ for \tz{} defined
 in the new \zexp{} parameterization definition.
This provides a simple to execute and well-defined procedure
 for converting \zexp{} results from one set of coefficients to another,
 allowing for direct comparison of \zexp{} coefficients
 even for independently defined parameterizations.

\subsection{Unknown Covariances}
\label{sec:unknown_covariances}

One challenge of combining \lqcd{} results is the presence of hidden correlations
 due to the reuse of gauge configurations and unreported fit covariances.
More than one procedure can be used to estimate the effects of such correlations,
 but these choices will only estimate the effects rather than definitively
 determine the contributions from correlations.
This subsection describes the methods used to estimate the effects
 of unknown correlations.
\sct{sec:covariance_derating} applies a technique called covariance
 derating to probe the full space of available covariances
 to construct an upper-bound.
As a comparison point, an alternative approach
 using the Schmelling procedure is detailed in
 \sct{sec:covariance_schmelling}.
\newcommand{\flag}{\text{FLAG}}%
This choice is akin to what is used
 by the Flavor Lattice Averaging Group (\flag)
 in other works involving \lqcd{}
 results~\cite{FlavourLatticeAveragingGroupFLAG:2024oxs}.

\subsubsection{Covariance Derating}
\label{sec:covariance_derating}

One method for addressing missing covariances is to
 use covariance derating~\cite{Koch:2024tit},
 which is particularly useful for the fit to form factor derivatives
 in \sct{sec:fitmethod_derivatives}.
This method starts from a ``naive'' covariance where all unknown correlations are set to 0.
After obtaining a best-fit result, all possible variations
 of the covariance matrix are tested up to a specified confidence level.
This yields a derating factor, which is multiplied into the covariance matrix
 to inflate the parameter uncertainties enough to cover the full range
 of possible variations.
For testing goodness-of-fit, another derating factor may be obtained
 that scales down the $\chi^{2}$ in a \pvalue{} computation.
The full functionality of the derating is available
 in the \nustattools~\cite{nustattools} package.

Some detail is required on the inputs to the derating algorithm.
The derating algorithm takes the naive covariance $C^{\bm{\alpha}}$
 of the input parameters $\bm{\alpha}$,
 the optimized fit parameters $\bm{a}$,
 and the Jacobian matrix $\bar{J}$ of the model prediction with respect
 to the best-fit parameters at the minimum,
\begin{align}
 \bar{J}_{ij}
 &= \frac{\partial r_{i}(\bm{a};\bm{\alpha})}{\partial a_{j}} .
\end{align}
The naive covariance is defined for the \zexp{} coefficients
 from \lqcd{} results in this work.
However, the fit ``data'' are not the \zexp{} coefficients themselves,
 but instead the form factor central values and derivatives
 computed from these coefficients.
A second Jacobian transformation brings the covariance
 over the \zexp{} coefficients to one over the form factor derivatives.
The transformation is
\begin{align}
 C^{F_{A}} &=
 J C^{\bm{\alpha}} J^{T} ,
 \nonumber\\
 J_{jk} &= \frac{r_{j}(\bm{a};\bm{\alpha})}{\partial \alpha_{k}} .
 \label{eq:jacobian_covariance}
\end{align}
The composed Jacobian transformation $J^{-1}\bar{J}$ is supplied as an input
 into the algorithm when computing the derating factors.

The total number of fit parameters does not always match the
 number of fit derivatives assumed in these fits.
This produces a rectangular Jacobian matrix in \eqn{eq:jacobian_covariance},
 which renders the composed Jacobian $J^{-1}\bar{J}$ uninvertible.
To circumvent this issue, dummy residuals $\tilde{r}$ are added to the fit
 and new dummy parameters $\tilde{a}$ are added to exactly reproduce those residuals.
This procedure does not change the total $\chi^{2}$ or the degree-of-freedom counting.
However, the presence of the dummy parameter enlarges the covariance,
 allowing for exploration of additional freedom from marginalized parameter combinations
 and yielding an invertible composed Jacobian.
In this work, the extra linear combinations are proportional to the
 \lqcd{} coefficients, i.e.
\begin{align}
 \tilde{r}_{i} = \alpha_{i} - \tilde{a}_{i} ,
\end{align}
 choosing the $\alpha_{i}$ to be the highest-order \zexp{} coefficients.

The derating factors $\beta$ indicate how much inflation is needed
 to account for unknown data correlations.
For assessing the uncertainties from fitting,
 a covariance derating factor $\beta^{\cov}$
 is obtained for scaling the covariance,
 enlarging the uncertainty:
\begin{align}
 C^{\derated} = \beta^{\cov} C^{\rm fit}.
 \label{eq:derating_covariance}
\end{align}
\newcommand{\gof}{\text{GoF}}%
\newcommand{\dof}{\text{DoF}}%
For assessing goodness-of-fit (\gof), a separate derating factor is computed
 that scales the $\chi^{2}$ from a fit with the naive covariance
 in a \pvalue{} computation with $\nu$ degrees of freedom (\dof):
\begin{align}
 \p^{\derated} = \p( \chi^{2} / \beta^{\gof}, \nu).
\end{align}
This derated \pvalue{} then serves as an upper bound on
 the possible agreement between models,
 allowing for tests of incompatibility
 but not guaranteeing compatibility.
In this work, both the derated \pvalue{}
 and the \pvalue{} obtained from a naive covariance matrix,
\begin{align}
 \p^{\naive} = \p( \chi^{2}, \nu),
\end{align}
 are reported.
The \pvalue{} computed from the naive covariance
 is not to be interpreted as a lower bound,
 instead only as an additional qualitative reference point.

\subsubsection{Schmelling Procedure}
\label{sec:covariance_schmelling}

An alternative approach for assessing missing covariances
 is the procedure outlined by Schmelling~\cite{Schmelling:1994pz}
 a choice that is commonly employed in \flag{}
 averages\footnote{%
 For more details, the reader is referred to Section~2.3
 of \rfr{FlavourLatticeAveragingGroupFLAG:2024oxs}.}.
For this procedure, two sampled quantities $x_{i}$ with unknown correlations
 are treated as fractionally correlated with a fraction $f$ in the range $[0,1]$,
 with a covariance matrix computed from
 the respective variances $\sigma_{i}^{2}$ as
\begin{align}
 C\big( x_{1}, x_{2}; f \big) &=
 \left( \begin{array}{cc}
 \sigma_{1}^{2} & f\sigma_{1}\sigma_{2} \\
 f\sigma_{1}\sigma_{2} & \sigma_{2}^{2}
 \end{array} \right)
 \nonumber\\
 &=
 \left( \begin{array}{cc}
 \sigma_{1} & 0 \\
 0 & \sigma_{2}
 \end{array} \right)
 \left( \begin{array}{cc}
 1 & f \\
 f & 1
 \end{array} \right)
 \left( \begin{array}{cc}
 \sigma_{1} & 0 \\
 0 & \sigma_{2}
 \end{array} \right) .
 \label{eq:schmelling_covariance}
\end{align}
The fraction $f$ ideally would be adjusted to yield a reduced $\chi^{2}$ of 1.

In \lqcd{} results, usually only the statistical uncertainty is correlated.
When this is true, using the full uncertainty in the offdiagonal correlations
 is not necessary.
The offdiagonal covariance can instead be modified to the value
\begin{align}
 f{\rm Cov}\big[x_{1}, x_{2}\big] &=
 f \sigma_{1;2} \sigma_{2;1}
 \nonumber\\
 \sigma_{i;j} &= \sqrt{ \sum_{\alpha} \big[ \sigma_{i}^{\alpha} \big]^{2} }
 \label{eq:schmelling_offdiagonal_only}
\end{align}
 where the sum inside the square root runs over only uncertainties
 that are correlated between the two results.
For this work, only the statistical uncertainty is assumed to be correlated
 between any two \lqcd{} results.
The only results in this work that exhibit these unknown correlations also
 separately report their statistical and systematic (or total) uncertainties,
 so no alternative procedures are considered.

\section{Summary of \lqcd{} Parameterizations}
\label{sec:lqcd_result_summary}

\lqcd{} results that will be included in fits are discussed in this section.
In particular, only \lqcd{} results that report complete error budgets,
 including chiral, continuum, and finite volume corrections,
 are subject to inclusion in the fits as part of this work.
However, several other collaborations have previous or ongoing efforts
 to further understand the form factors with \lqcd{}~\cite{
 Meyer:2016kwb,
 Hasan:2017wwt,
 Ishikawa:2018jee,
 Shintani:2018ozy,
 Lin:2020wko,
 He:2021yvm,
 Ishikawa:2021eut,
 Meyer:2021vfq,
 Ohta:2021ldu,
 Barca:2022uhi,
 Ohta:2022csu,
 Koponen:2022gfe,
 Ohta:2023ygq,
 Hackl:2024whw,
 Aoki:2025taf,
 Tsuji:2025quu,
 Barone:2025hhf}.
The parameterizations used by each collaboration are given
 in detail in \scts~\ref{sec:rqcd}--\ref{sec:pndme}.
The results that are used are compared in \sct{sec:lqcd_comparison}.

\subsection{\rqcd{} 2020}
\label{sec:rqcd}

The \rqcd{} parameterization is defined as a $\kmax=2$ \zexp{}
 with no sum rule constraints imposed.
The covariance for the \rqcd{} result is not available in the literature.
However, the supplemental material for ~\rfr{RQCD:2019jai}
 contains an array of means and errors for a large range of $Q^{2}$ values.
These values can be used to reverse engineer the parameterization and covariances
 used to generate those values good up to at least 4 significant digits.
From this reverse engineering, the \rqcd{} parameterization is found to be
\begin{align}
 \big( a_{0}, a_{1}, a_{2} \big)
 &= \big( 1.01333032, -1.71328189, -0.59048148 \big)
\end{align}
 with statistics-only covariance
\begin{align}
 \left( \begin{array}{lll}
 \phm0.00055973 & -0.00192868 & \phm0.00075550 \\
 -0.00192868 &  \phm0.01415581 & -0.02897714 \\
 \phm0.00075550 & -0.02897714 & \phm0.09590674
 \end{array} \right)
\end{align}
 and a covariance for statistics and systematics together,
\begin{align}
 \left( \begin{array}{lll}
 \phm0.00088551 & -0.00468337 & \phm0.01000718 \\
 -0.00468337 & \phm0.08434121 & -0.16947322 \\
 \phm0.01000718 & -0.16947322 & \phm0.35842788
 \end{array} \right) .
\end{align}
The reverse engineered parameterization yields the values
 $\tc=0.17150090~\GeVsq$ and $\tz=-0.17147395~\GeVsq$.

\subsection{\nme{} 2022}
\label{sec:nme}

The \nme{} parameterization is given as a \zexp{} parameterization
 up to $\kmax=2$ with no sum rules imposed.
The \zexp{} parameters are given in Eq. (56) of \rfr{Park:2021ypf} as
\begin{align}
 ( a_{0}, a_{1}, a_{2} )
 &= \Big( 0.725(5), -1.63(3), -0.17(13) \Big)
\end{align}
 with $\tc=(3\cdot 0.135)^{2}~\GeVsq$ and $\tz=-0.50~\GeVsq$.
The covariance that accompanies these parameters is listed in Eq. (I2) as
\begin{align}
 \left( \begin{array}{lll}
 \phm2.188\times10^{-5} &    -2.238\times10^{-5} &    -1.155\times10^{-4} \\
    -2.238\times10^{-5} & \phm8.549\times10^{-4} & \phm2.769\times10^{-3} \\
    -1.155\times10^{-4} & \phm2.769\times10^{-3} & \phm1.811\times10^{-2}
 \end{array} \right) .
\end{align}

Though \nme{} tests the effects of systematic shifts due
 to the chiral, continuum, and finite volume extrapolations,
 the reported covariance does not include uncertainties for this variations.
This leads to a form factor with precision below 1\% for the entire $Q^{2}$ range,
 below the uncertainty due to isospin-breaking corrections to the
 form factor~\cite{Cirigliano:2022hob}.
\nme{} notes that these extrapolations do not change the central value significantly
 and that allowing for these variations increases the uncertainty
 by about a factor of 3~\cite{park:privatecomm}.

To allow for uncertainty due to the extrapolation to the physical point,
 the \nme{} covariance is scaled up by a factor of $3^{2}=9$ or $5^{2}=25$
 for the purposes of this work.
Even with the larger scale factor of 25, the \nme{} result
 is the most precise and drives the uncertainty of the final average.
Use of the smaller scale factor reduces the uncertainty by as much
 as 30\% over the larger scale factor.
For the remainder of this work, the larger scale factor of 25 is used
 to give a more conservative estimate and to allow the other \lqcd{}
 results more contribution to the final average.

\subsection{\mainz{} 2022}
\label{sec:mainz}

The \mainz{} result~\cite{Djukanovic:2022wru}
 defines a $\kmax=2$ parameterization
 with $\tc = (3\cdot 0.135)^{2}~\GeVsq$
 and $\tz = 0$.
The \zexp{} coefficients are
\begin{align}
 ( a_{0}, a_{1}, a_{2} )
 = \Big( &\;\,\phm1.225(\phz39)(\phz25),
 \nonumber\\ &-1.274(237)(\phz70),
 \nonumber\\ &-0.379(592)(179) \Big)
\end{align}
 where the first uncertainty ($\sigma^{\stat}$) is statistical
 and the second ($\sigma^{\syst}$) is systematic.
The total error is computed by summing the errors in quadrature,
\begin{align}
 \sigma^{\total}_{i} &= \sqrt{\sigma^{\stat}_{i} +\sigma^{\syst}_{i} } .
\end{align}
With the given correlation matrix
\begin{align}
 \rho &=
 \left( \begin{array}{rrr}
   1\phantom{.00000}, & -0.67758, &  0.61681 \\
  -0.67758, &  1\phantom{.00000}, & -0.91219 \\
   0.61681, & -0.91219, &  1\phantom{.00000} \\
 \end{array} \right) ,
\end{align}
 the covariance matrix is then computed from
\begin{align}
 \Sigma^{\total}_{ij} = \sigma^{\total}_{i} \rho_{ij} \sigma^{\total}_{j} .
 \label{eq:covariance_from_correlation}
\end{align}

\subsection{\etm{} 2023}
\label{sec:etm}

The \etm{} result~\cite{Alexandrou:2023qbg} defines a $\kmax=3$ parameterization
 with $\tc = (3\cdot 0.135)^{2}~\GeVsq$
 and $\tz = 0$.
The \zexp{} coefficients are
\begin{align}
 &( a_{0}, a_{1}, a_{2}, a_{3} )
 \nonumber\\
 &= \Big( 1.245(28), -1.19(18), -0.54(61), -0.1(1.3) \Big)
\end{align}
 where the total uncertainty is given.
The correlation matrix
\begin{align}
 \rho &=
 \left( \begin{array}{rrrr}
   1\phantom{.000}, & -0.421, & \phm0.247, & -0.246 \\
  -0.421, &  1\phantom{.000}, & -0.918, & \phm0.799 \\
  \phm0.247, & -0.918, &  1\phantom{.000}, & -0.952 \\
  -0.246, & \phm0.799, & -0.952, & 1\phantom{.000} \\
 \end{array} \right) ,
\end{align}
 is given, where the covariance matrix is computed
 with \eqn{eq:covariance_from_correlation}.

\subsection{\pndme{} 2023}
\label{sec:pndme}

The \pndme{} result~\cite{Jang:2023zts}
 defines a $\kmax=2$ parameterization
 with $\tc = (3\cdot 0.140)^{2}~\GeVsq$
 and $\tz = -0.25~\GeVsq$.
The \zexp{} coefficients are
\begin{align}
 ( a_{0}, a_{1}, a_{2} )
 = \Big( 0.876(28),
 -1.669(99),
 0.483(498) \Big)
\end{align}
 where the total uncertainty is given.
The correlation matrix
\begin{align}
 \rho &=
 \left( \begin{array}{rrr}
   1\phantom{.00000}, & -0.45170, & -0.02966 \\
  -0.45170, &  1\phantom{.00000}, & -0.24394 \\
  -0.02966, & -0.24394, &  1\phantom{.00000} \\
 \end{array} \right)
\end{align}
 is given, where the covariance matrix is again computed
 with \eqn{eq:covariance_from_correlation}.

\subsection{\lqcd{} Comparisons}
\label{sec:lqcd_comparison}

The \lqcd{} results obtained by different collaborations
 are nontrivial checks of consistency with each other.
The strongest cross-validation of these results is the
 use of different quark actions, which are governed by
 different effective theories under an expansion in the lattice spacing.
Strictly speaking, \lqcd{} results with different actions cannot
 be compared with each other before taking the continuum limit,
 as they will have unknown corrections due to
 lattice spacing discretization errors.
Additionally, results with different quark action necessarily
 use independently-generated gauge field configurations
 with different correlation functions.
Results with nonoverlapping sets of gauge field configurations are
 therefore uncorrelated with each other by construction.

\lqcd{} relies on computations across several ensembles of gauge configurations.
These gauge configurations have a variety of
 masses, lattice spacings, and volumes to control systematic
 effects from finite simulations.
Physical mass ensembles are the gold standard,
 although they are more computationally costly than unphysical ensembles
 and have larger variances that lead to larger statistical uncertainties.
Using unphysically heavy mass ensembles allows for more precise
 explorations of systematics due to lattice spacing and finite volume effects.
The extrapolation to physical mass typically
 relies on Heavy Baryon Chiral Perturbation Theory~\cite{
 Jenkins:1990jv,
 Jenkins:1991ts},
 which does not describe the extrapolation of baryon quantities
 to physical mass well~\cite{
 Walker-Loud:2008rui,
 PACS-CS:2009cvn}.
As a result, performing computations on one or more physical mass ensembles
 to anchor the extrapolation endpoint is desirable.

Each collaboration also has their own unique choice of fit methodology.
After generating correlation functions as a function of Euclidean time,
 the correlation functions must be fit to extract the corresponding
 energies and matrix elements.
The matrix elements are fit as a function of 4-momentum transfer $Q^{2}$
 to some fit parameterization, then the parameters of these form factors
 are extrapolated to the physical point
 at the chiral, continuum, and infinite-volume limits.
These are the parameterizations reported as the form factors for QCD.
The process for addressing these steps is not unique
 and different approaches confer different advantages and drawbacks.

Many collaborations fit to the known Euclidean time-dependence
 of the form factors using Bayesian priors,
 which can be written as (neglecting terms related to the finite volume):
\begin{align}
 C^{2pt}(t) &\approx \sum_{k=1}^{\infty} |z_{k}|^{2} e^{-E_{k}t} ,
 \nonumber\\
 C^{3pt}(t,\tau;{\cal J}) &\approx \sum_{k,\ell=1}^{\infty}
 z_{k} z_{\ell}^{\ast} {\cal J}_{k\ell} e^{-E'_{k}(t-\tau)} e^{-E_{\ell}\tau} .
 \label{eq:lqcd_correlation_functions}
\end{align}
\newcommand{\twopt}{\text{two-point}}%
\newcommand{\thrpt}{\text{three-point}}%
Here, $C^{2pt}$ and $C^{3pt}$ refer to the \twopt{} and \thrpt{} functions, respectively.
The \twopt{} functions give access to overlap factors $z_{k}$ and state energies $E_{k}$.
The overlap factors connect the vacuum to states $|k\rangle$
 with the desired quantum numbers through an
 ``interpolating operator'' ${\cal O}$,
\begin{align}
 z_{k} = \langle 0 | {\cal O} | k \rangle .
 \label{eq:overlap_factor}
\end{align}
The index $k$ connects to the ground state ($k=1$)
 as well as higher excited states ($k\geq2$),
 with $k=0$ being reserved for the vacuum.
In practice, the number of excited states included in the fits
 is limited to a finite value.
Contamination from improperly-constrained excited states
 is a concern in final results and typically part of the
 systematic error budget.

Given the overlap factors and energies computed with \twopt{} functions,
 the \thrpt{} functions additionally give access to matrix elements of current ${\cal J}$,
\begin{align}
 {\cal J}_{k\ell} = \langle k | {\cal J} | \ell \rangle .
 \label{eq:matrix_element}
\end{align}
To extract the axial matrix elements,
 the interpolating operator ${\cal O}$
 is typically a three-quark operator
 with some Fourier phase to impart a distinct 3-momentum,
 which couples strongly to a nucleon.
The states $|k\rangle$ that appear in
 \eqns~(\ref{eq:overlap_factor})~and~(\ref{eq:matrix_element})
 would then be states with the same quantum numbers as the nucleon,
 including excitations such as nucleon-pion scattering states,
 and the current insertion ${\cal J}$
 could be defined as an axial current
 with fixed 3-momentum transfer $\bm{q} = \bm{p}'-\bm{p}$,
\begin{align}
 {\cal J}_{k\ell} \to \langle N(\bm{p}') |
 \sum_{\bm{x}} {\cal A}(\bm{x}) e^{-i\bm{q}\cdot\bm{x}}| N(\bm{p}) \rangle .
\end{align}

As an alternative to fitting the correlation functions to the known forms
 in \eqn{eq:lqcd_correlation_functions},
 \rqcd{} uses a form that captures contributions from
 nucleon-pion excited states.
In nucleon axial matrix elements,
 excited state terms with nucleon-pion states are typically
 difficult to constrain with \twopt{} functions alone
 due to volume-suppressed overlap factors
 but are enhanced in \thrpt{} functions due to their coupling to amplitudes.
The form used by \rqcd{} adds terms with energies fixed to the values
 $E, E'\to E_{\pi} + E_{N}$ and provides overlaps mimicking
 the kinematic dependence expected from chiral perturbation theory.
This allows the fitter limited freedom with which to constrain
 nucleon-pion states without getting lost in parameter space.
For the explicit form, the reader is referred to \eqns~(2.44)--(2.47)
 of \rfr{RQCD:2019jai}.

The other alternative fit employed by \mainz{} is the summation method~\cite{
 Maiani:1987by,
 Gusken:1989qx,
 Dong:1995rx,
 Capitani:2010sg}.
This technique works by summing over the time dependence of the
 \thrpt{} correlation functions, which yields
\begin{align}
 R(t) = \sum_{\tau=a}^{t-a}& C^{3pt}(t,\tau) /\sqrt{C^{2pt}_{\bm{p}'}(t) C^{2pt}_{\bm{p}}(t)}
 \nonumber\\
 &\underset{t\gg a}{\longrightarrow} \; {\rm constant} + t {\cal J}_{11} + O(t e^{-\Delta E t}) ,
 \label{eq:summation_method}
\end{align}
 where $\Delta E$ is the gap in energies between the ground state
 and the first excited state.
The ground state matrix element is then extracted
 by fitting the slope of $R$ with respect to $t$.
This benefits from increased suppression of excited states
 and reduced dependence on choice of fit parameters.

The summary of results used in this work are listed in \tbl{tab:lqcd_comparison}.
The number of gauge ensembles used,
 and the size of the subset of those ensembles at physical masses,
 are listed in the third and fourth columns.
For the number of ensembles, higher counts typically
 lead to better control of systematics.

The three separate quark actions used in these results
 are listed in the fifth column of \tbl{tab:lqcd_comparison}.
The results from \rqcd{} and \mainz{} share subsets of the same
 gauge ensembles from the Coordinated Lattice Simulation (CLS) effort~\cite{Bruno:2014jqa},
 with nonperturbatively order $a$-improved Wilson clover fermions~\cite{
 Sheikholeslami:1985ij,
 Bulava:2013cta},
 and generate their own correlation functions on those ensembles.
Sharing gauge ensembles leads to correlations in the results
 which are unknown and will need to be addressed in this work.
Though the \nme{} result employs the same fermion action as \rqcd{} and \mainz,
 the gauge ensembles used by \nme{} are generated independently and are uncorrelated.
\newcommand{\hisq}{\text{HISQ}}
The result from \pndme{} also uses a Wilson clover action for the ``valence'' quarks
 (those quarks that are explicitly included in the correlation function constructions),
 these correlation functions are computed on gauge ensembles
 that were generated with a Highly-Improved Staggered Quark (\hisq) action~\cite{
 Follana:2006rc,
 Hao:2007iz,
 Hart:2008sq}
 for the quarks in the ``sea'' (virtual quark-antiquark loops).
These results are therefore also uncorrelated with the other Wilson clover results.
Lastly, the \etm{} result uses the twisted mass action~\cite{
 Frezzotti:2000nk,
 Frezzotti:2003ni,
 ETM:2010iwh},
 which is also uncorrelated with the others.

The fit methods, which were outlined in this section,
 are enumerated in the sixth column of \tbl{tab:lqcd_comparison}.
In broad terms, the three methods used for fitting the correlation
 functions to extract energies and matrix elements
 are the ``Bayesian prior exponential'' method,
 which involves fitting to the forms in \eqn{eq:lqcd_correlation_functions} directly
 with priored coefficients,
 or the $\chi$PT-inspired and summation (\eqn{eq:summation_method}) method techniques employed by
 \rqcd{} and \mainz{}, respectively.

The last column of \tbl{tab:lqcd_comparison} gives the parameterizations
 from the \lqcd{} results that were used in this work.
The employed parameterizations are all \zexp.
The order of the expansions are listed,
 with the \zexp{} either being carried up to $\kmax=2$ (labeled as $z^{2}$)
 or $\kmax=3$ ($z^{3}$).
For all of the \zexp{} parameterizations, no sum rules were used.
Since $g_{A}$ is an output of \lqcd{} computations,
 no constraint fixing the intercept was employed either.

\newcommand{\nosumrules}{no sum rules}
\begin{table*}
\begin{tabular}{llrrllr}
Collaboration & Ref. &
$N_{\rm ens}$ & $N_{\rm ens}^{\rm phys}$ & quark action & fit method & parameterization
\\
\hline
\rqcd{} 2020 & \cite{RQCD:2019jai}
& 37 & 2 & clover & $\chi$PT-inspired & $z^2$, \nosumrules
\\
\nme{} 2022 & \cite{Park:2021ypf}
&  7 & 0 & clover & Bayesian prior exponential & $z^2$, \nosumrules
\\
\mainz{} 2022 & \cite{Djukanovic:2022wru}
& 14 & 1 & clover & summation
 & $z^2$, \nosumrules
\\
\pndme{} 2023 & \cite{Jang:2023zts}
& 13 & 2 & clover (on HISQ) & Bayesian prior exponential & $z^2$, \nosumrules
\\
\etm{} 2023 & \cite{Alexandrou:2023qbg}
& 3 & 3 & twisted mass & Bayesian prior exponential & $z^3$, \nosumrules
\end{tabular}\\[3em]
\caption{
 Comparison of the relevant details of the \lqcd{} results used in this work.
 The first and second columns give the name of the collaboration
 and the reference for the work.
 The third and fourth columns respectively denote the total number of ensembles
 and the number of ensembles at physical pion mass for each work.
 The quark action is given in the fifth column.
 The \pndme{} result uses a different quark action for the valence (Wilson clover)
 and sea quarks (Highly-Improved Staggered Quarks (\hisq)).
 The fit method used to extract excited states and axial matrix elements
 is given in the sixth column.
 Finally, the parameterization reported as a final result
 in each reference is given in the last column.
 \label{tab:lqcd_comparison}
}
\end{table*}

\section{Fit Results}
\label{sec:fit_results}

This section delves into the results of fitting to the \lqcd{} form factors.
\sct{sec:fits_derivatives} explores the fits to form factor derivatives
 outlined in \sct{sec:fitmethod_derivatives}.
\sct{sec:fits_formfactor} likewise goes into detail about sampled form factors
 outlined in \sct{sec:fitmethod_sampled}.
Several systematic variations are employed to test the sensitivity
 to various choices.
\scts~\ref{sec:fits_derivatives}~and~\ref{sec:fits_formfactor}
 deal with systematics that are specific to each of the fit methods,
 respectively.
\sct{sec:fits_comparison} deals with systematics that are common
 to both methods.

Throughout this section,
 the plotted form factors are normalized by the deuterium result
 in \rfr{Meyer:2016oeg} to improve the visibility of the curves.
Uncertainties using the derivatives fit method of \sct{sec:fits_derivatives}
 are inflated by the derating factor $\beta^{\cov}$
 according to \eqn{eq:derating_covariance}
 to account for unknown correlations between \lqcd{} results.
Unlike the sister paper~\cite{Meyer:inprep},
 no regularization term is used for the \zexp{} coefficients
 in this work, instead relying on the residuals to constrain
 the form factor values.
No exceptionally large parameter values are observed in any of the fit results.

\subsection{Fits to Derivatives}
\label{sec:fits_derivatives}

In this subsection,
 the fits employing analytic derivatives of the form factors
 to construct residuals are explored.
The philosophy behind choosing this fit method is that the
 form factor should be insensitive to higher-order \zexp{} coefficients,
 where empirically the slope and intercept are enough to constrain the form factor.
If the fits are optimized over the central value and first derivative of the form factor
 at only a single value of $Q^{2}$, the form factor shape is still
 described well over a reasonably large range of $Q^{2}$.
Adding evaluations at more $Q^{2}$ points does not provide much additional constraint
 on the shape beyond what is obtained from linear behavior with $z$.

In the present work, the $Q^{2}$ behavior of the form factor is expected
 to be described well over the range 0--1~\GeVsq, governed by the range of $Q^{2}$
 that is probed by \lqcd{} computations.
This range can be extended if \lqcd{} computations are carried out
 for $Q^{2}$ values larger than 1~\GeVsq{}.

Much of the behavior in the range 1--2~\GeVsq{}
 in this work is controlled by the sum rules in \eqn{eq:sum_rules}.
Without imposing these sum rules, there is no expectation
 that the form factor should be well behaved at all.
Although it is more theoretically justified to forgo sum rules
 to avoid potentially adding bias, including the sum rules is a practical consideration
 for experimental applications in Monte Carlo event generators where it is necessary
 to statistically sample the form factor out to large $Q^{2}$.
The figures in the remainder of the work are plotted out to 2~\GeVsq{}
 so that the extrapolation of the form factor beyond the expected 1~\GeVsq{}
 cutoff can be examined qualitatively.
The effect of removing the sum rules is also explored
 in more detail in \sct{sec:fitting_sumrules}.

The resulting fit quality for all of the fits to derivatives that appear in this subsection
 are listed in \tbl{tab:results_fit_derivatives}.
The first column gives the description of the fits involved.
These descriptions include at least the \kmax{} and the highest order
 of derivatives used as residuals (either 1st derivatives or 2nd derivatives).
Where appropriate, alternative values of \Qtmin{} and \Qtcom,
 as defined in \eqn{eq:q2comparison},
 or \Qtanc, the $Q^{2}$ value for an included anchor point,
 are also listed.

The naive $\chi^{2}$, computed assuming all unknown correlations are set to 0,
 and \dof{} are listed for each fit in \tbl{tab:results_fit_derivatives}.
In addition, the derating factors from \sct{sec:covariance_derating}
 and the corresponding \pvalue{s} are given.
The value $\p^{\derated}$ can be treated as an upper bound on the \pvalue{}
 given the variations in the covariance matrix.
If $\p^{\derated}$ is less than 0.05,
 this can be used as evidence to exclude those fit assumptions.
However, neither \pvalue{} represents a lower bound:
 even if both \pvalue{s} are greater than 0.05,
 this does not guarantee that the corresponding fit model is consistent
 with the chosen \lqcd{} inputs.
The last two columns give the first two fit parameters that are optimized in the fits.
In the case of \zexp{} fits,
 the $a_{1}$ coefficient
 and $a_{2}$ coefficient (when $\kmax>5$ or no sum rules are used)
 are reported.
The dipole fit lists the value for $M_{A}$ that is obtained from the fit.

\begin{table*}[htb!]
 \begin{tabular}{l|rrrrrrrr}
 Description
 & $\chi^{2}$
 & $\nu$
 & $\beta^{\gof}$
 & $\beta^{\cov}$
 & $\p^{\naive}$
 & $\p^{\derated}$
 & $a_{1}\pha$
 & $a_{2}\pha$
 \\
 \hline
$\kmax=5$, 1st derivatives
& $21.11$ & $ 9$ & 1.19 & 1.21 & 0.01 & 0.04 & $-1.901(\phz\phz8) \pha$ & ---\pha            \\
$\kmax=6$, 1st derivatives (nominal)
& $ 7.14$ & $ 8$ & 1.20 & 1.18 & 0.52 & 0.65 & $-1.701(\phz42)    \pha$ & $\phm 0.26(\phz9)  \pha$ \\
$\kmax=7$, 1st derivatives
& $ 7.00$ & $ 7$ & 1.19 & 1.16 & 0.43 & 0.55 & $-1.658(\phz95)    \pha$ & $\phm 0.43(35)     \pha$ \\
$\kmax=6$, 2nd derivatives
& $25.35$ & $13$ & 1.17 & 1.21 & 0.02 & 0.06 & $-1.638(\phz11)    \pha$ & $\phm 0.02(\phz5)  \pha$ \\
$\kmax=7$, 2nd derivatives
& $13.73$ & $12$ & 1.00 & 1.16 & 0.32 & 0.32 & $-1.741(\phz25)    \pha$ & $-0.05(\phz5)      \pha$ \\
$\kmax=8$, 2nd derivatives
& $10.13$ & $11$ & 1.14 & 1.12 & 0.52 & 0.63 & $-1.563(\phz74)    \pha$ & $\phm 0.94(39)     \pha$ \\
$\kmax=6$, 1st derivatives, $\Qtcom=0.25~\GeVsq$
& $12.11$ & $ 8$ & 1.24 & 1.18 & 0.15 & 0.28 & $-1.535(\phz37)    \pha$ & $\phm 0.07(\phz7)  \pha$ \\
$\kmax=6$, 1st derivatives, $\Qtcom=0.50~\GeVsq$
& $12.55$ & $ 8$ & 1.23 & 1.22 & 0.13 & 0.25 & $-1.646(\phz44)    \pha$ & $\phm 0.18(\phz9)  \pha$ \\
$\kmax=6$, 1st derivatives, $\Qtmin=0.00~\GeVsq$
& $ 5.50$ & $ 8$ & 1.05 & 1.05 & 0.70 & 0.73 & $-1.687(\phz30)    \pha$ & $\phm 0.29(\phz8)  \pha$ \\
$\kmax=6$, 1st derivatives, $\Qtmin=0.10~\GeVsq$
& $11.05$ & $ 8$ & 1.23 & 1.19 & 0.20 & 0.34 & $-1.660(\phz51)    \pha$ & $\phm 0.15(\phz9)  \pha$ \\
$\kmax=7$, 1st derivatives, $\Qtanc=0.10~\GeVsq$
& $15.63$ & $12$ & 1.17 & 1.13 & 0.21 & 0.35 & $-1.777(\phz29)    \pha$ & $-0.14(\phz5)      \pha$ \\
$\kmax=7$, 1st derivatives, $\Qtanc=0.75~\GeVsq$
& $20.82$ & $12$ & 1.28 & 1.10 & 0.05 & 0.17 & $-1.734(\phz27)    \pha$ & $-0.20(15)         \pha$ \\
$\kmax=7$, 1st derivatives, $\Qtanc=1.00~\GeVsq$
& $21.60$ & $12$ & 1.01 & 1.12 & 0.04 & 0.05 & $-1.756(\phz30)    \pha$ & $-0.21(16)         \pha$ \\
$\kmax=2$, 1st derivatives, no sum rules
& $ 8.16$ & $ 8$ & 1.23 & 1.18 & 0.42 & 0.58 & $-1.726(\phz77)    \pha$ & $-0.27(20)         \pha$ \\
$\kmax=3$, 1st derivatives, no sum rules
& $ 6.76$ & $ 7$ & 1.20 & 1.16 & 0.45 & 0.58 & $-1.620(102)       \pha$ & $\phm 0.87(76)     \pha$ \\
$\kmax=6$, 1st derivatives, $t_{0}=-0.25$ GeV$^{2}$
& $ 7.51$ & $ 8$ & 1.19 & 1.18 & 0.48 & 0.61 & $-1.771(\phz36)^{\ast}$ & $0.35(\phz8)^{\ast}$ \\ 
\hline
\hline
 Description
 & $\chi^{2}$
 & $\nu$
 & $\beta^{\gof}$
 & $\beta^{\cov}$
 & $\p^{\naive}$
 & $\p^{\derated}$
 & $M_{A}\pha$ &\\
\hline
dipole, 1st derivatives
& $13.41$ & $ 9$ & 1.14 & 1.19 & 0.14 & 0.23 & $ 1.219(\phz15)   \pha$ & ---\pha                  \\
 \end{tabular}
\caption{
 Summary of fit results for fits to derivatives,
 enumerated in the same order that they first appear
 in figures within this section.
 The first column gives a description of the fit that was computed.
 The naive $\chi^{2}$,
 assuming all of the unknown covariances are 0,
 is given in the second column.
 The third column, labeled $\nu$, gives the \dof{} for the fit.
 The fourth and fifth columns give the derating factors
 $\beta^{\gof}$ and $\beta^{\cov}$, respectively,
 as discussed in \sct{sec:covariance_derating}.
 The \pvalue{} computed using the naive $\chi^{2}$ is listed in the sixth column,
 and the derated \pvalue{} accounting for all variations
 of the covariance matrix within a 99\% confidence interval
 is given in the seventh column.
 The second-to-last column gives the \zexp{} coefficient multiplying the $z^{1}$ term
 in the power series, which is the lowest-order fit parameter,
 or the value of $M_{A}$ for the dipole fit.
 The last column gives the coefficient multiplying the $z^{2}$ term for the \zexp{} fits.
 For the row labeled with a description
 ``$\kmax=6$, 1st derivatives, $t_{0}=-0.25$ GeV$^{2}$,''
 marked with an asterisk,
 the printed \zexp{} coefficients $a_{1}$ and $a_{2}$
 are the translated coefficients using the prescription set out in
 \sct{sec:zexp_translated}.
 The raw \zexp{} coefficient values obtained for this fit are
 $(a_{1}, a_{2}) = (-1.671(48), -0.68(9))$.
\label{tab:results_fit_derivatives}
}
\end{table*}

\subsubsection{Included Derivatives}

The fits explored in this subsection test the sensitivity
 to where the form factor values and derivatives are evaluated.
This includes the number of evaluation points,
 at what $Q^{2}$ values they are probed,
 and what quantities are probed at each point.

The nominal choice of where to evaluate the form factors and derivatives
 is at the value $Q^{2}=-\tz$,
 using the \tz{} defined for each individual \lqcd{} result.
This is the expansion point for the fit \zexp{} parameterization
 and the most natural place to evaluate derivatives for that parameterization,
 as discussed in \sct{sec:fitmethod_derivatives}.
In cases where $\tz=0$,
 the comparison point is moved slightly away from $Q^{2}=0$.
The default point where evaluations for each \lqcd{} result are done
 is summarized by the value
\begin{align}
 \Qtcom = {\rm max} \big[ \Qtmin,\, |-\tz| \big] ,
 \label{eq:q2comparison}
\end{align}
 with $\Qtmin=0.05~\GeVsq$ as the nominal choice.
Other choices for \Qtmin{} will be explored in \sct{sec:derivatives_q2evaluation}.

The nominal fit takes only the central value and first derivative
 of the form factor for each \lqcd{} result,
 although the process generalizes to higher-order derivatives.
This provides 2 residuals per \lqcd{} result,
 both the central value and derivative evaluated at \Qtcom.
The parameterizations have at least 3 \zexp{} coefficients though,
 so there is still some freedom to select other residuals to fit against.
The most natural option is to include second-order derivatives at the same
 evaluation point.
Another option is to add an anchor point,
 the form factor central value evaluated at some higher $Q^{2}$ point.

\begin{figure}[htb!]
\includegraphics[width=0.48\textwidth]{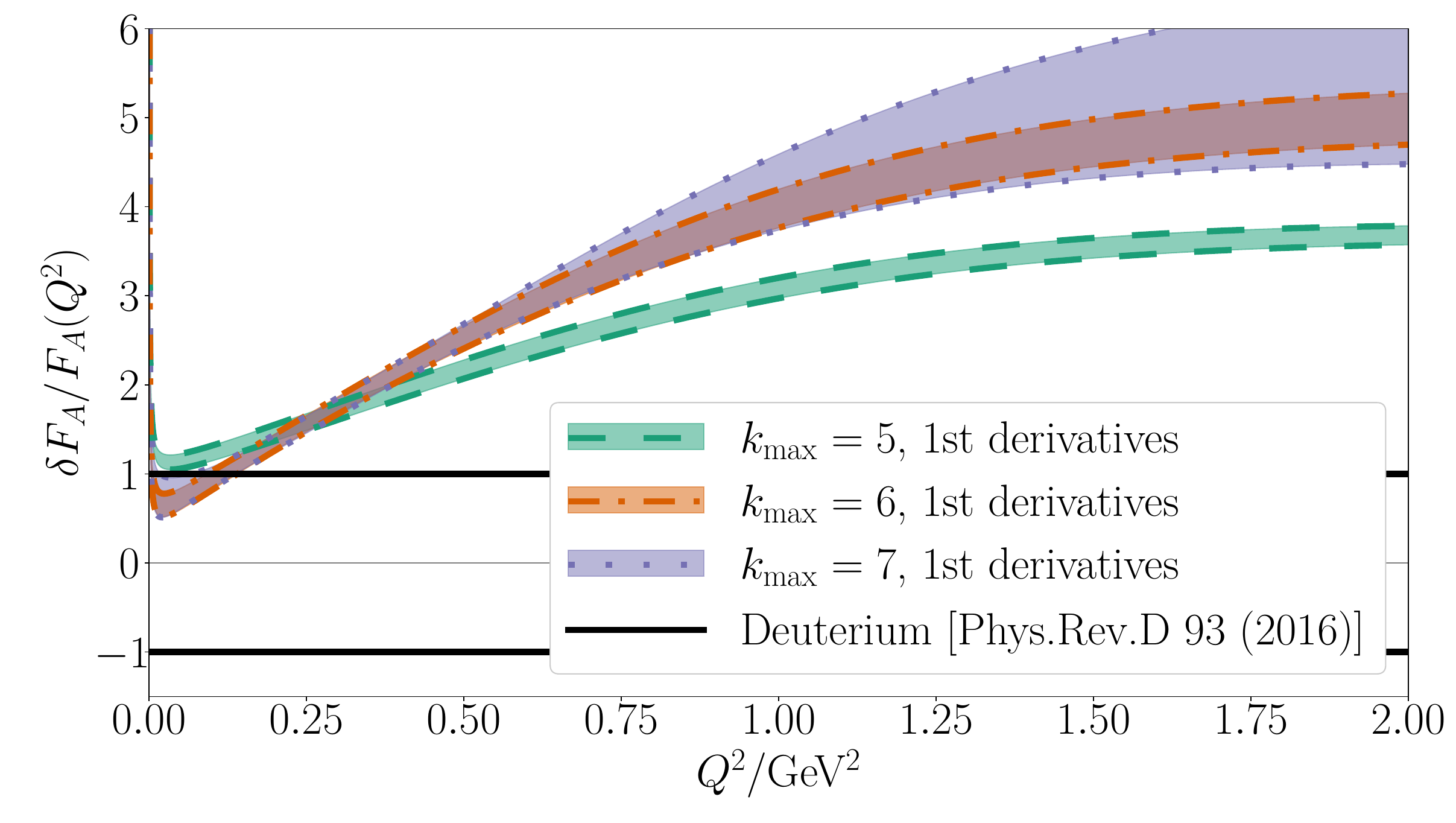}
\caption{
 Plot of the axial form factor fits from this work
 normalized by the \zexp{} deuterium result from \rfr{Meyer:2016oeg}.
 The uncertainty band for the fit with $\kmax=5$
 and up to first-order derivatives only is given by
 the teal shaded region bordered by dashed lines.
 A similar fit with $\kmax=6$ ($\kmax=7$) is shown
 as the orange band with a dot-dashed border
 (blue-violet band with a dotted border).
 For reference, the normalized range $0\pm1$ of the \zexp{}
 result is shown as black lines.
\label{fig:derivatives_kmax}
}
\end{figure}

A comparison of fits to first derivatives only with
 different choices of $\kmax$ is shown in \fgr{fig:derivatives_kmax}.
The nominal fit with $\kmax=6$ is shown against $\kmax=5$ and $\kmax=7$ fits.
The $\kmax=5$ result shows indications of being overfit,
 with a poor goodness-of-fit and lower \pvalue{s}.
The $\kmax=5$ fit also disagrees with the $\kmax=6$ fit across the majority
 of the $Q^{2}$ range.
The reported value of $\p^{\derated}$ for $\kmax=5$ is just below the threshold
 such that this fit ansatz can be ruled out.
The $\kmax=7$ fit has a similar goodness-of-fit and \pvalue{s} to the $\kmax=6$ fit,
 but a larger uncertainty at high $Q^{2}$.
The decrease of $\chi^{2}$ resulting from increasing \kmax{} to 7
 does not appreciably improve the \pvalue, suggesting that the
 additional parameter only acts to inflate the uncertainties without providing
 a substantially better description of the input data.
Since the $\kmax=7$ fit includes only up to first-order derivatives,
 the only reason there is any constraint on the $a_{3}$ coefficient
 is because the $\Qtcom$ is offset to $\Qtmin$ for the parameterizations
 expanded at $\tz=0$.
The fit residuals are still mainly dominated by the contributions
 from the coefficients $a_{1}$ and $a_{2}$,
 so the constraint on $a_{3}$ is not substantial.
This is reflected in the fit posterior $a_{3} \approx 1.06(97)$,
 with effectively 100\% uncertainty.

\begin{figure}[htb!]
\includegraphics[width=0.48\textwidth]{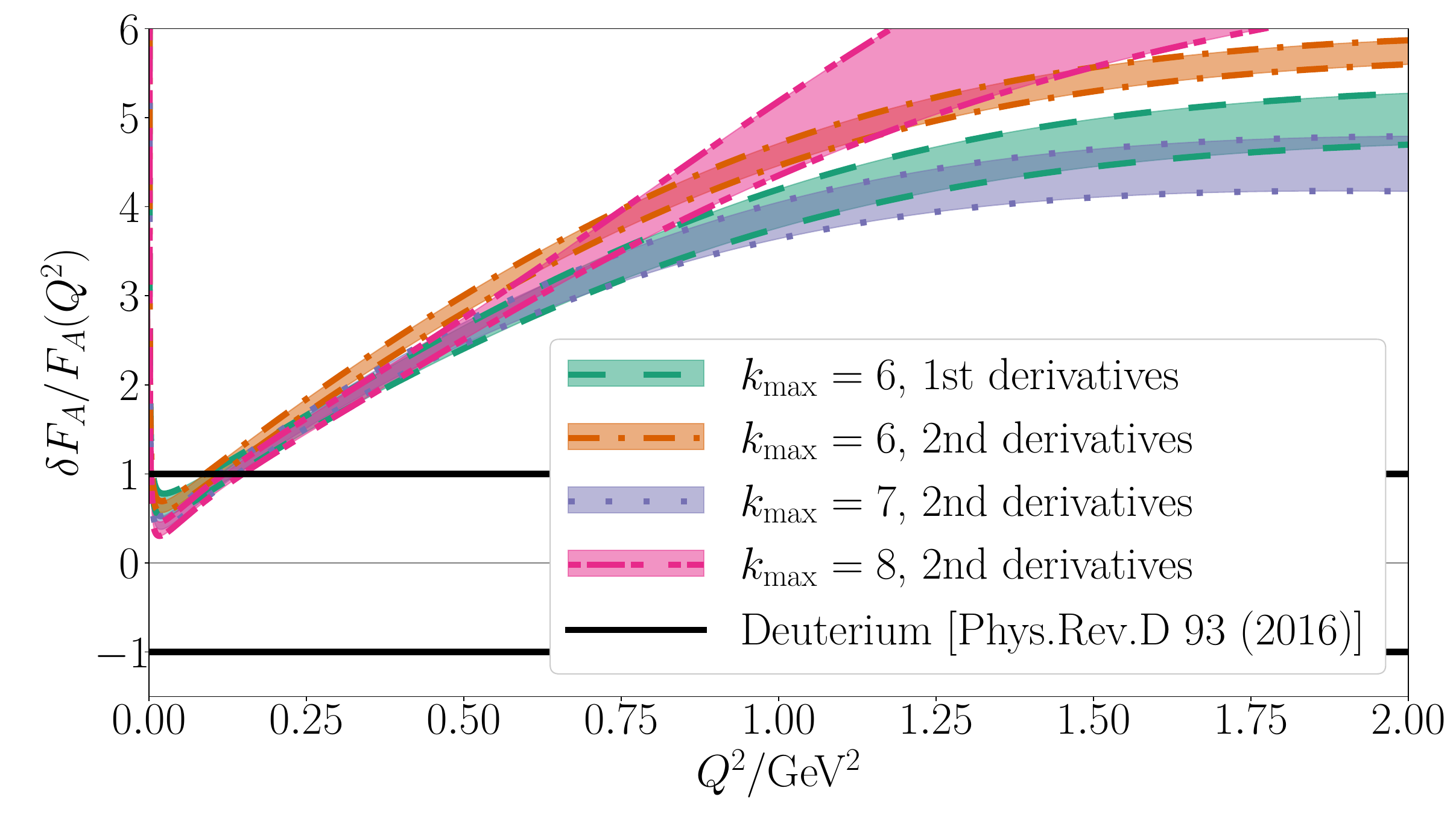}
\caption{
 Plot of the axial form factor fits from this work
 normalized by the \zexp{} deuterium result from \rfr{Meyer:2016oeg}.
 The uncertainty band for the nominal result with $\kmax=6$
 and up to first-order derivatives only is given by
 the teal shaded region bordered by dashed lines.
 Fits with up to second derivatives for $\kmax=6$, 7, and 8
 are shown as
 an orange band with a dot-dashed border,
 a blue-violet band with a dotted border,
 and a pink band with a triple-dash border,
 respectively.
 For reference, the normalized range $0\pm1$ of the \zexp{}
 result is shown as black lines.
\label{fig:2ndderivatives_kmax}
}
\end{figure}

A comparison of fits with up to second derivatives for
 different choices of $\kmax$ is shown in \fgr{fig:2ndderivatives_kmax}.
This plot also shows the nominal fit, $\kmax=6$ with up to first derivatives,
 for reference.
To account for the extra constraints on the fit parameters,
 $\kmax$ is chosen to be 1 larger than in \fgr{fig:derivatives_kmax}.
Like with the fits to first derivatives,
 the $\kmax=7$ fit to second derivatives is largely identical to the nominal
 fit in the range $Q^{2}\sim0.1$--$1.0~\GeVsq$.
For $Q^{2}\lesssim0.1~\GeVsq$, the fit is still compatible within $2\sigma$.
The $\kmax=6$ fit to second derivatives departs from the nominal fit,
 accompanied by poor \pvalue{s}.
The smallest \kmax{} ($\kmax=6$ here) again gives low \pvalue{s}
 that are small enough to be deemed incompatible,
 but the $\kmax=7$ and 8 fits have \pvalue{s} that are acceptable.
The $\kmax=8$ fit has a compatible shape for $Q^{2}\lesssim0.5~\GeVsq$,
 but then increases to values considerably larger than the nominal fit.

Increasing to $\kmax=8$ yields a \pvalue{} comparable to the nominal fit,
 suggesting that the additional degrees of freedom
 are meaningfully constrained by the fit parameters.
These coefficients are again only probed by the residuals
 for the $\tz=0$ results with $\Qtmin\neq0$,
 as was the case for $\kmax=7$ fits with up to first derivatives
 in the discussion of \fgr{fig:derivatives_kmax}.
However, the parameters fixed by sum rule constraints in this fit take
 absolute values larger than 20.
By contrast, the $\kmax=7$ fit with second derivatives
 has sum rule parameters up to absolute values of less than 6,
 and the nominal fit with $\kmax=6$ has parameter absolute values less than 2.
If a regularization term was included in this fit,
 the regularization would add a significant penalty to the $\chi^{2}$
 for the $\kmax=8$ fit, spoiling the apparent compatibility of the fit.
For this reason, the $\kmax=8$ fit is not considered a good fit.

\subsubsection{\texorpdfstring{$Q^{2}$}{Momentum Transfer} Evaluation Points}
\label{sec:derivatives_q2evaluation}

Modifications of the specific values of $Q^{2}$ where the residuals
 are evaluated are considered as another alternative method for
 evaluating sensitivity to systematics.
Testing for compatibility between these different choices
 of included residuals gives confidence in the selection
 of a nominal fit.
There are a number of selections of $Q^{2}$ that are considered throughout
 the derivative fitting process:
\begin{enumerate}
 \item the $Q^{2}$ where the residuals for the primary evaluations are performed,
 labeled by \Qtcom;
 \item the $Q^{2}$ for primary evaluations that are expanded about 0,
 more specific than \Qtcom{} and labeled by \Qtmin{} in \eqn{eq:q2comparison}; and
 \item the $Q^{2}$ for any additional anchor points,
 labeled by \Qtanc.
\end{enumerate}
These three possibilities will be tested in this subsection.

\fgr{fig:derivatives_q2compare} shows the results of
 testing sensitivity to the value of \Qtcom.
Three choices all at $\kmax=6$ are utilized,
 either evaluating derivatives at the nominal choice defined in \eqn{eq:q2comparison}
 or fixing all residuals to be evaluated at one choice of $Q^{2}$.
The value $\Qtcom=0.25~\GeVsq$ is chosen to approximate
 $-\tz=0.28~\GeVsq$ used in \rfr{Meyer:2016oeg},
 and $-\tz=0.50~\GeVsq$ as the natural expansion point
 for the nominal fit in this work.
While the fits employing \eqn{eq:q2comparison} are constructed in a way
 to be dominated by only the lowest-order \zexp{} coefficients,
 as indicated in \eqn{eq:fit_matching_condition},
 the fits with fixed \Qtcom{} produce a nonzero $z$ and
 probe all of the \zexp{} coefficients at once.

Qualitatively, the fits in \fgr{fig:derivatives_q2compare}
 appear largely consistent with each other,
 with some slight tensions at the 2--$3\sigma$ level between the nominal choice
 using \eqn{eq:q2comparison} and the $\Qtcom=0.25~\GeVsq$ expansion point.
These tensions are mainly confined to $Q^{2}$ below about $0.2~\GeVsq$
 and $Q^{2}$ above around $0.75~\GeVsq$.
The values of $a_{1}$ obtained for this fit is in tension as well,
 with the $\Qtcom=0.25~\GeVsq$ giving a $3.0\sigma$ difference.
In contrast, the $\Qtcom=0.50~\GeVsq$ fit
 exhibits much better agreement across the board and
 only a $0.9\sigma$ tension between values of $a_{1}$.
The tensions appear to come from disagreements between
 the \lqcd{} results on the slope at the new $Q^{2}$ points.
Although the tensions are present in both fits with fixed \Qtcom,
 the residuals tend to pull equally in both directions for $\Qtcom=0.50~\GeVsq$,
 whereas the pull is biased toward smaller $a_{1}$ in the $\Qtcom=0.25~\GeVsq$ fit.
The fixed expansion points come with mildly decreased goodness-of-fit,
 yielding \pvalue{s} less than 0.3, compared to the $\p^{\derated}$
 bound of 0.65 for the nominal fit.
This suggests that the fixed expansion point is less favorable
 than the nominal choice, although not appreciably so.

\begin{figure}[htb!]
\includegraphics[width=0.48\textwidth]{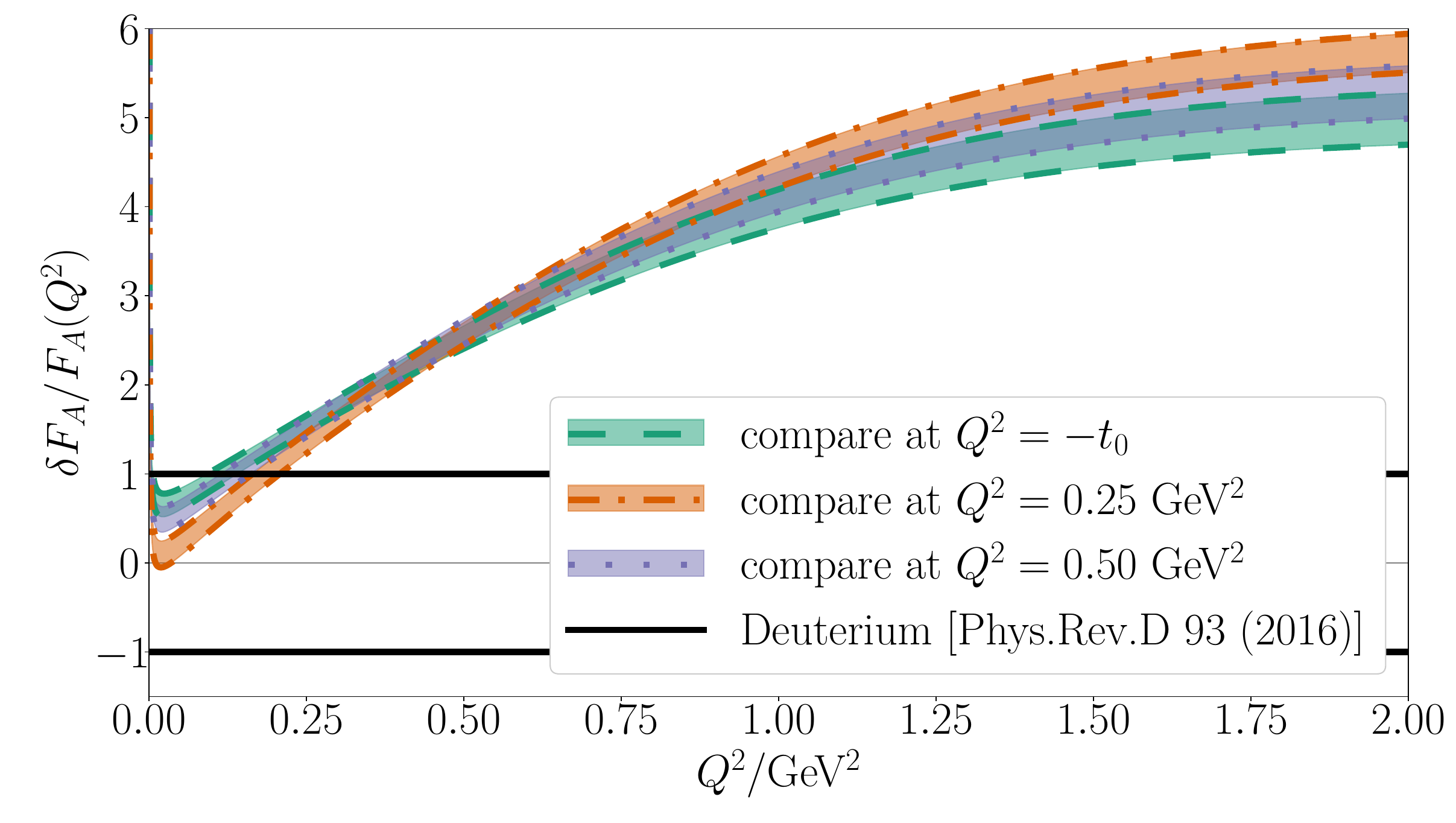}
\caption{
 Plot of the axial form factor fits from this work
 normalized by the \zexp{} deuterium result from \rfr{Meyer:2016oeg}.
 The uncertainty band for the nominal result,
 evaluating the form factor and its derivatives for
 each separate \lqcd{} result at its own value of $Q^{2}=-\tz$,
 is given as the teal shaded region bordered by dashed lines.
 When $\tz=0$, the minimum value $Q^{2}_{0}$ is used in place of $\tz$,
 as described in the text.
 Fits evaluating the form factor and its derivatives
 at $Q^{2}=0.25~\GeVsq$~and~$0.50~\GeVsq$
 are shown as the orange band with a dot-dashed border
 and the blue-violet band with a dotted border, respectively.
 For reference, the normalized range $0\pm1$ of the \zexp{}
 result is shown as black lines.
\label{fig:derivatives_q2compare}
}
\end{figure}

Rather than evaluating all of the \lqcd{} results to a single $Q^{2}$,
 a less disruptive choice is to only change the value of \Qtmin{}
 used in \eqn{eq:q2comparison}.
The nominal choice is $\Qtmin=0.05~\GeVsq$,
 but the other choices $\Qtmin=0$ and $\Qtmin=0.10~\GeVsq$
 are also explored in \fgr{fig:derivatives_q20}.

Choosing $\Qtmin=0$ is entirely consistent with the nominal choice,
 including a consistent value of $a_{1}$,
 smaller uncertainty, and even slightly better \pvalue{s}.
However, this choice ignores the complication that arises from
 fixing the value of $g_{A}$, as mentioned in \sct{sec:fitmethod_derivatives}.
For this reason, the ansatz with $\Qtmin=0$ is not chosen as a nominal fit.
The other choice at $\Qtmin=0.10~\GeVsq$ is also consistent with the nominal choice,
 with less than a $1\sigma$ shift in $a_{1}$ and the form factor values
 across the entire plotted $Q^{2}$ range.
The \pvalue{s} are also lower than that of the nominal fit,
 but still at a reasonable level of $\p^{\derated}\lesssim0.4$.

\begin{figure}[htb!]
\includegraphics[width=0.48\textwidth]{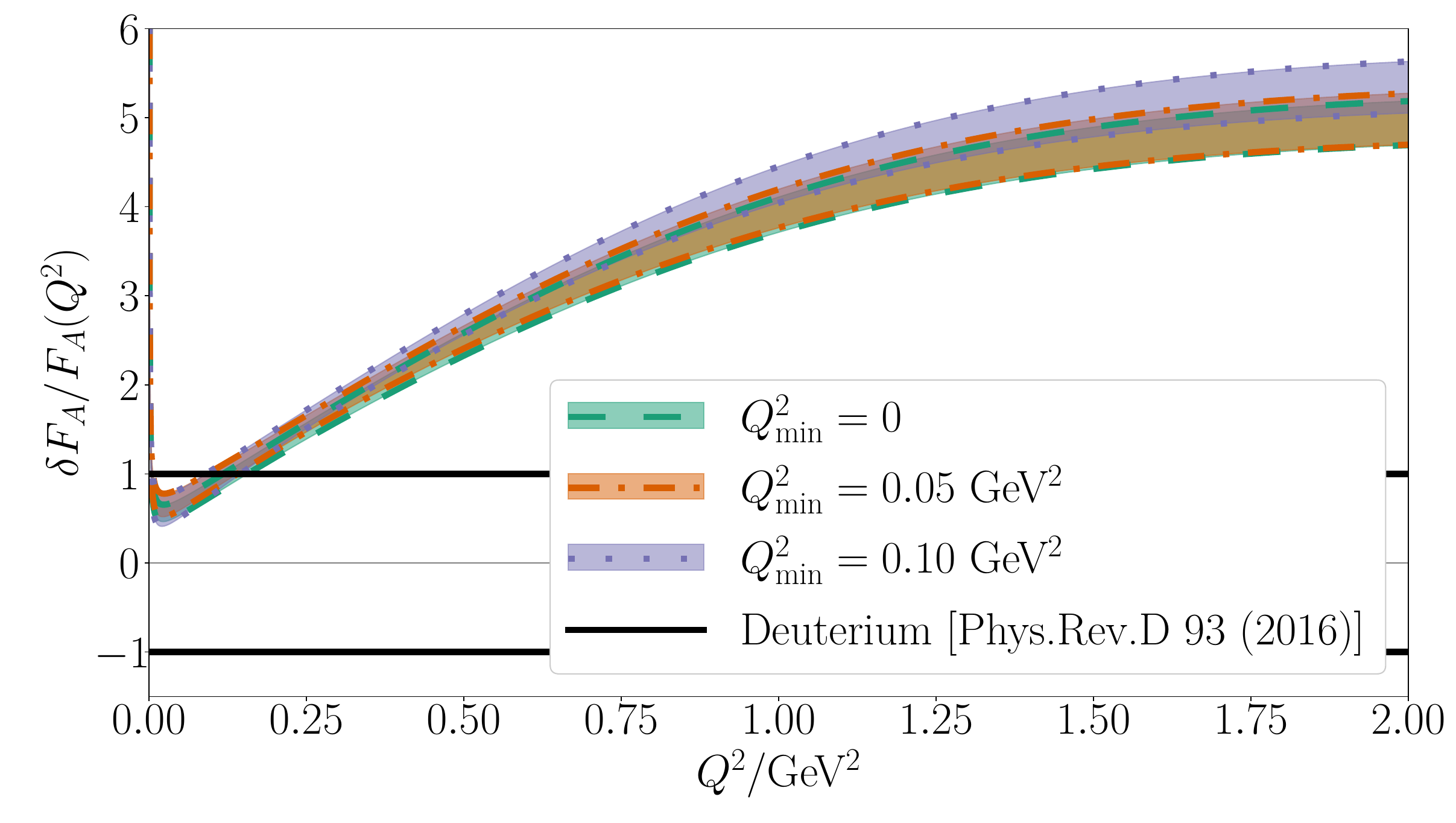}
\caption{
 Plot of the axial form factor fits from this work
 normalized by the \zexp{} deuterium result from \rfr{Meyer:2016oeg}.
 The teal dashed, orange dot-dashed, and blue-violet dotted
 regions correspond to setting the minimum $Q^{2}$ evaluation point,
 denoted by \Qtmin, equal to $0$, $0.05~\GeVsq$, and $0.10~\GeVsq$, respectively.
 This value is used for evaluation of the form factors and derivatives
 for results that set $\tz=0$ (\etm{} and \mainz).
 The nominal fit corresponds to the choice $0.05~\GeVsq$,
 given by the orange dot-dashed region.
 For reference, the normalized range $0\pm1$ of the \zexp{}
 result is shown as black lines.
\label{fig:derivatives_q20}
}
\end{figure}

\fgr{fig:derivatives_anchor} shows the nominal $\kmax=6$ fit compared
 to $\kmax=7$ fits that include an anchor point at various values of \Qtanc.
The choices \Qtanc$=0.10~\GeVsq$, which sits between the $Q^{2}=-\tz$ evaluation points,
 as well as $0.75$ and $1.00~\GeVsq$ beyond the largest value of \tz{} are shown for comparison.
All of the choices yield
 similar fits across the range $Q^{2}\sim0.1$--$1.25~\GeVsq$
 when compared to each other and the nominal fit.
Only mild tensions $\lesssim1.5\sigma$ are observed for $a_{1}$ across the fits.
The \pvalue{s} are reduced when including an anchor point,
 mostly due to growing tensions at larger values of $Q^{2}$,
 and only the fit with $\Qtanc=1.00~\GeVsq$ has a \pvalue{}
 low enough to be deemed unacceptable.

\begin{figure}[htb!]
\includegraphics[width=0.48\textwidth]{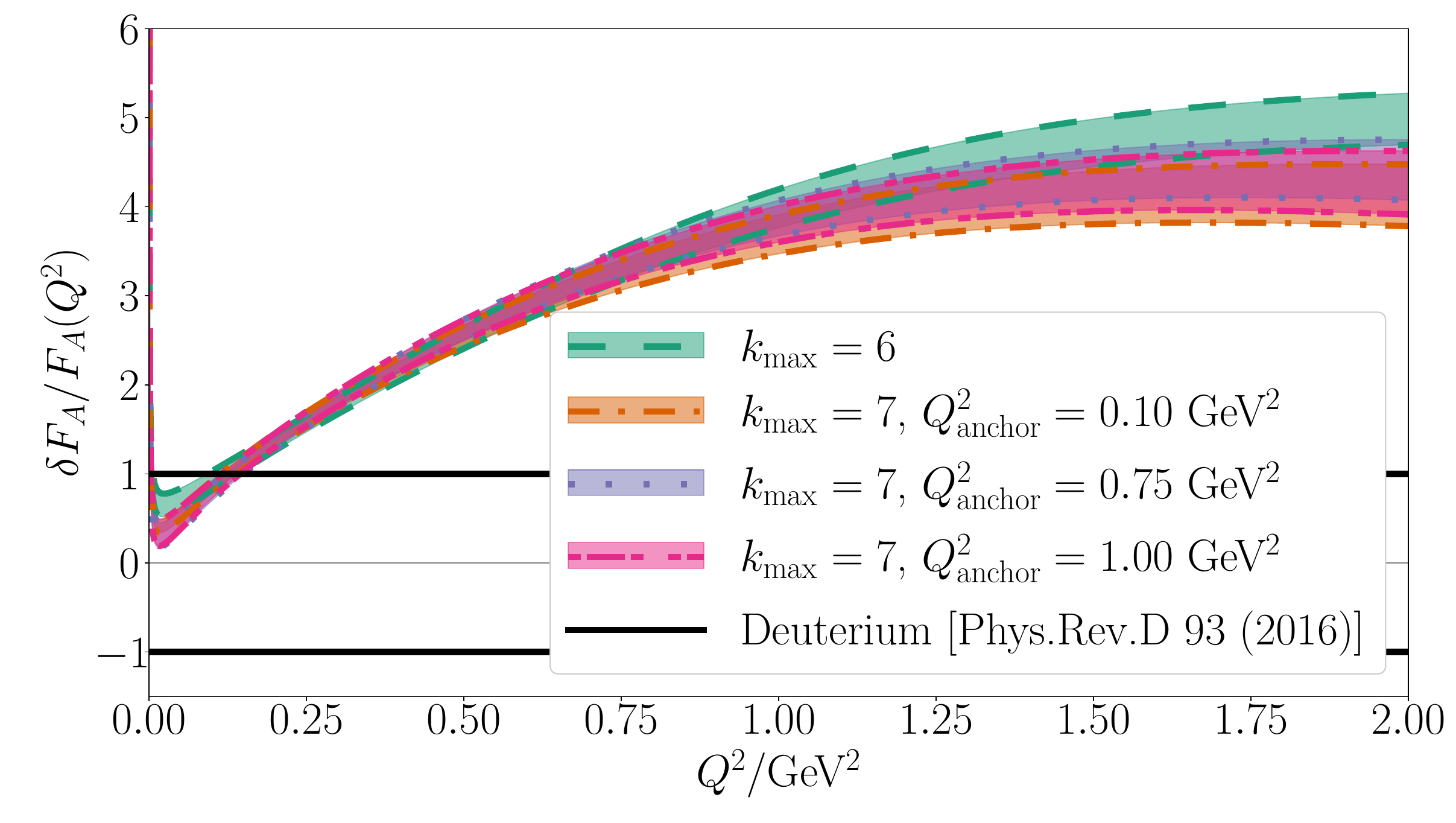}
\caption{
 Plot of the axial form factor fits from this work
 normalized by the \zexp{} deuterium result from \rfr{Meyer:2016oeg}.
 The teal dashed region corresponds to the nominal fit with no anchor point.
 The orange dot-dashed, blue-violet dotted, and pink triple-dashed ranges,
 which all lie on top of each other,
 correspond to $\kmax=7$ fits with an anchor point
 at $\Qtanc=0.10$, $0.75$, and $1.00~\GeVsq$, respectively.
 For reference, the normalized range $0\pm1$ of the \zexp{}
 result is shown as black lines.
\label{fig:derivatives_anchor}
}
\end{figure}

\subsection{Fits to Sampled Form Factors}
\label{sec:fits_formfactor}

Fits to the form factor over a range of equidistant $Q^{2}$ values
 are also considered as a more conventional alternative to fitting
 derivatives in the previous subsection.
Like with the derivatives fits,
 form factors in this subsection are fit with sum rules
 and presented as ratios to the deuterium fit of \rfr{Meyer:2016oeg}.

These sampled fits are described in \sct{sec:fitmethod_sampled}.
A covariance matrix is assembled over the parameters
 from each of the \lqcd{} results that has been modified
 according to the Schmelling procedure in \sct{sec:covariance_schmelling}.
One thousand randomly sampled sets of parameters are produced
 respecting the known covariances.
These parameters are used to compute form factor central values
 for $Q^{2}$ in the range $\{ 0.0, 0.1, \dots, 1.0 \}~\GeVsq$
 for each \lqcd{} result.
Residuals and an uncorrelated $\chi^{2}$ loss function with
 a fixed set of errors are constructed according to \eqn{eq:residuals_sampled}.
Additional fits were performed testing with 4 times the random samples
 and 4 times smaller spacing between $Q^{2}$ values,
 both producing negligible difference compared to the nominal choice.

Unlike the derivatives fits, a proper goodness-of-fit
 cannot be assessed for these sampled form factor fits.
The goodness-of-fit can be made to produce a \pvalue{}
 arbitrarily close to 1 by fitting a denser set of $Q^{2}$.
However, comparisons of reduced $\chi^{2}$ values for a given
 fixed set of $Q^{2}$ can still be performed to assess
 the relative fit quality of similar sampled fits.

Joint fits with the sampled form factor and other data
 cannot be easily performed either,
 for example fitting the \lqcd{} results together
 with other neutrino scattering data.
Like with the \pvalue{s}, choosing a denser set of $Q^{2}$
 would correspondingly increase the number of residuals
 in an overall $\chi^{2}$.
The fit results could be arbitrarily skewed toward the \lqcd{}
 by adding more $Q^{2}$ evaluation points.
This is a primary reason to favor the derivatives fits to the form factor,
 since the weight of the \lqcd{} in the overall fit is more limited.

\subsubsection{Schmelling Factor Variation}

Fits with different Schmelling scale factors $f$
 are depicted in \fgr{fig:sampling_schmelling}.
One striking observation
 is that the uncertainty band for the fully correlated ($f=1$) fit
 is significantly larger than the uncertainty band of the nominal fit
 with the derivatives method.
These two methods handle the unknown covariances quite differently from each other.
The derivatives method makes use of the derating procedure,
 which more directly quantifies the amount of uncertainty inflation
 by exploring the full space of available correlated variations,
 allowing for variations where different offdiagonal elements take different
 relative sizes.
The form factor sampling method used here applies the Schmelling procedure,
 which scales all of the unknown covariances with the same factor $f$.

The three values $f=1$, 0.5, and 0, all fit with $\kmax=6$,
 are shown together.
The scale $f$ is only applied to the statistical uncertainty
 of the \lqcd{} parameters that have unknown cross-correlations.
These three fits have drastically different uncertainties,
 where $f=0$ (with unknown correlations set to exactly 0)
 corresponds to a form factor uncertainty about 50\% the size
 of the fully-correlated ($f=1$) result at $Q^{2}=0.5~\GeVsq$
 and 30\% at $Q^{2}=1.0~\GeVsq$.

Given the relative size of the uncertainties of the form factor method
 versus the derivatives method, the uncorrelated ($f=0$) case will be adopted
 as default for the remaining form factor method variations.
Note that a larger Schmelling factor will only inflate the uncertainties
 without changing the central value, so compatibility between fits
 will only get better as $f$ increases.
The direct comparison between the two fit methods with the
 uncorrelated and fully-correlated fits be revisited in \sct{sec:fits_comparison}.

\begin{figure}[tbh!]
\includegraphics[width=0.48\textwidth]{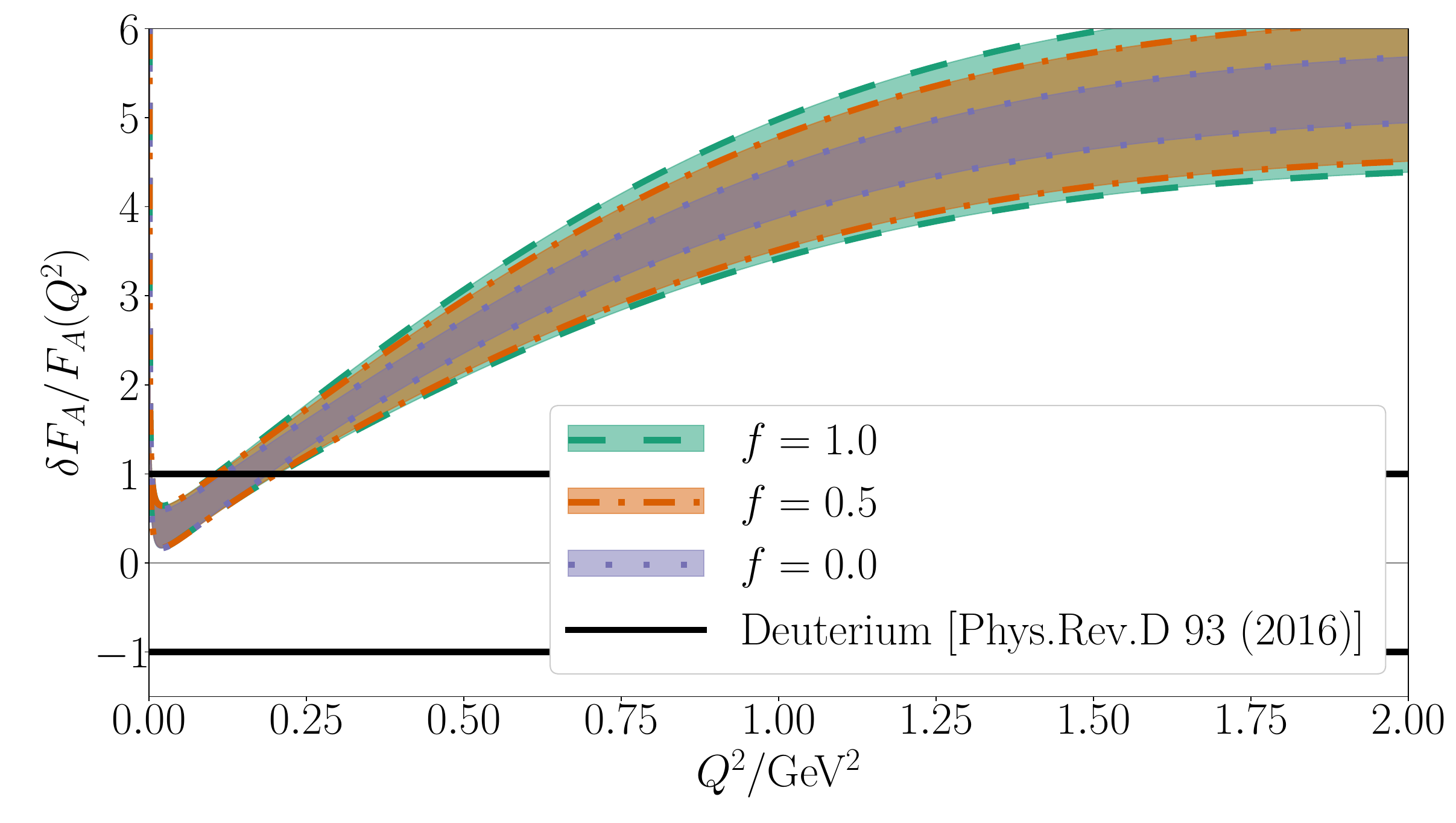}
\caption{
 Plot of the axial form factor fits from this work
 normalized by the \zexp{} deuterium result from \rfr{Meyer:2016oeg}.
 All fits shown here have $\kmax=6$.
 The teal dashed, orange dot-dashed, and blue-violet dotted regions
 correspond to the fits with the offdiagonal scale factor
 for the Schmelling procedure (defined in \eqn{eq:schmelling_covariance}) set to
 $f=1$, 0.5, and 0, respectively.
 For reference, the normalized range $0\pm1$ of the \zexp{}
 result is shown as black lines.
\label{fig:sampling_schmelling}
}
\end{figure}

\subsubsection{\texorpdfstring{\kmax}{Maximum expansion order} Variation}

The plot of variations of the fit form factor over increasing \kmax{}
 is shown in \fgr{fig:sampling_kmax}.
The corresponding reduced $\chi^{2}$ values for the
 $\kmax=5$, 6, and 7 fits are 1.49, 0.82, and 0.83, respectively.
Like for the derivatives fit, the $\kmax=5$ result is likely overfit
 and departs from other fit results at low and intermediate $Q^{2}$.
Increasing to $\kmax=7$ does not significantly improve over $\kmax=6$,
 yielding a larger uncertainty at high $Q^{2}$ and a slightly larger reduced $\chi^{2}$.
The fit parameter values for the $\kmax=6$ fit are (taking $a_{3}$ from the sum rules)
\begin{align}
 \big( a_{1}, a_{2}, a_{3} \big)
 = \Big( -1.627(63), 0.18(12), 1.11(37) \Big) ,
\end{align}
 compared to the fit parameters for the $\kmax=7$ fit,
\begin{align}
 \big( a_{1}, a_{2}, a_{3} \big)
 = \Big( -1.648(73), -0.06(28), 1.52(65) \Big) .
\end{align}
The fit parameters for these fits are unsurprisingly consistent with each other.

\begin{figure}[tbh!]
\includegraphics[width=0.48\textwidth]{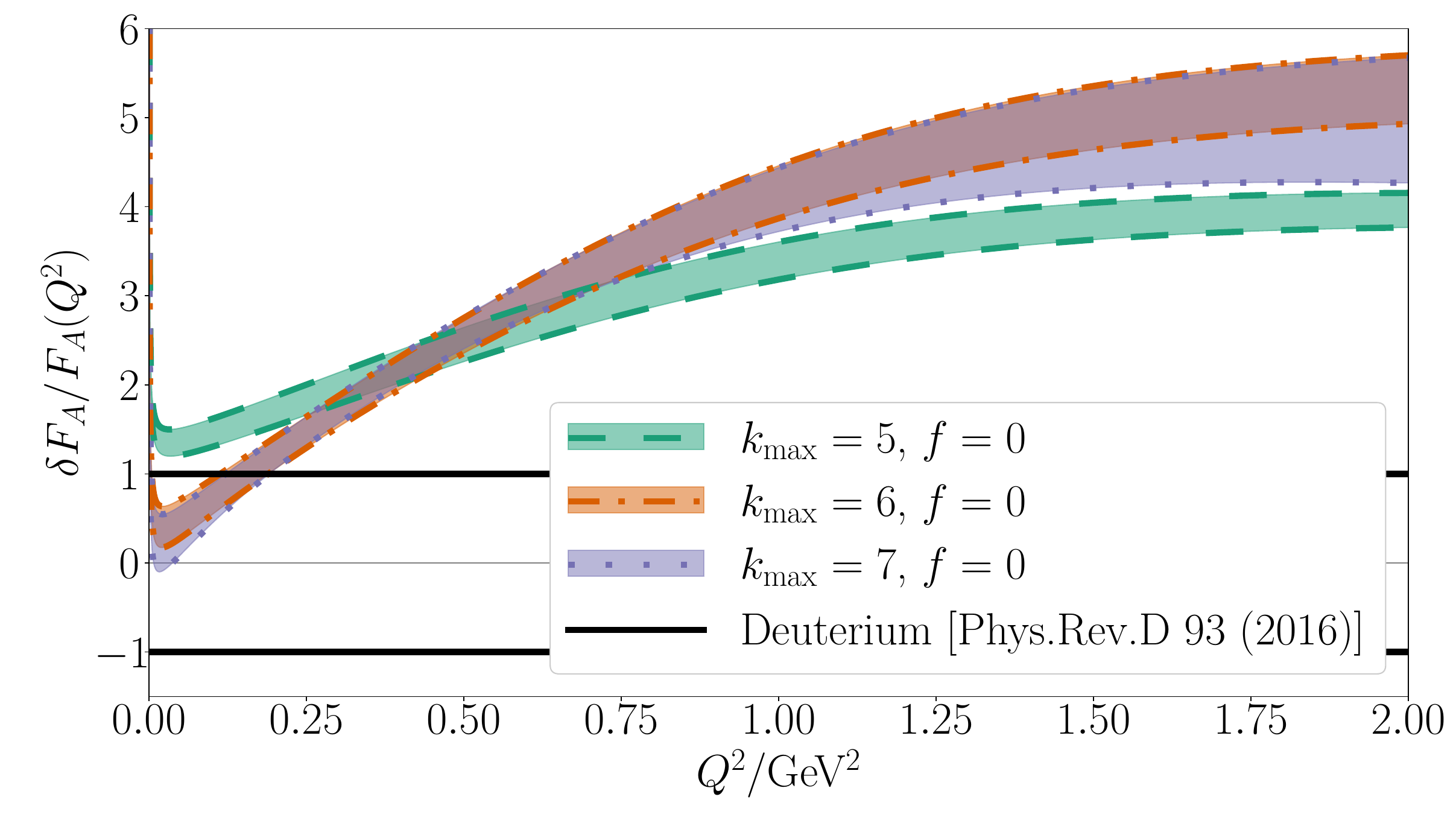}
\caption{
 Plot of the axial form factor fits from this work
 normalized by the \zexp{} deuterium result from \rfr{Meyer:2016oeg}.
 The teal dashed, orange dot-dashed, and blue-violet dotted regions
 correspond to the fits with $\kmax=5$, 6, and 7, respectively.
 For reference, the normalized range $0\pm1$ of the \zexp{}
 result is shown as black lines.
\label{fig:sampling_kmax}
}
\end{figure}

\subsection{Comparisons Between Methods}
\label{sec:fits_comparison}

Having explored a number of form factor systematics for both averaging methods
 to decide on the nominal fit choices, the results of the averaging schemes
 can now be compared to each other.
The method of fitting derivatives will be taken as the nominal choice for
 the final fit result, with the sampled form factor fit used as a validation.
There are a number of reasons to prefer the derivatives fit over
 the sampled form factor:
\begin{enumerate}
 \item the residuals for the derivatives fit can be defined analytically
 rather than stochastically;
 \item fitting to derivatives has a natural choice for selecting
 the $Q^{2}$ to compare against, specifically at $Q^{2}=-\tz$;
 \item when number of residuals is less than the rank of the covariance matrix,
 the transformation from the set of input set of \zexp{} parameters to the
 residuals preserves the invertibility of the covariance matrix
 without additional modifications.
\end{enumerate}
The analytic form of the derivatives method makes it simpler
 to form a combined fit with other sources,
 such as with experimental neutrino scattering data.

The comparison of the nominal choices for each fit method are shown
 in \fgr{fig:method_nominal}.
The fit to form factor derivatives is more precise
 than the sampled form factor fits even for the uncorrelated ($f=0$)
 form factor sampling method.
The relative size of the (correlated) derivatives fit with derating
 and the (uncorrelated) sampled fit might lead to the assumption
 that the \lqcd{} results are closer to uncorrelated than fully correlated.
However, correlations are not the only factor at play in the
 overall precision of the fits.
The differences between the assumptions of the two fits,
 namely only examining the central values versus also examining
 the first derivatives, can change the overall precision of the fit.
It is likely that both details play a role in the final fit precision.

The derivatives fit stays within the $1\sigma$ band
 of the uncorrelated sampled fit over nearly the entire $Q^{2}$ range,
 only departing for $Q^{2}\lesssim0.1~\GeVsq$.
This trend is also seen in \fgr{fig:derivatives_anchor},
 where the different choices of $Q^{2}$ anchor point
 drag the form factor below the nominal fit at low $Q^{2}$.
This could be a consequence of slight tensions between
 the parameterizations preferred from fitting the central values
 versus the first derivatives.
Adding the anchor points gives twice as many central values
 informing the fits as derivatives, which will bias the fit
 in favor of the central values of the \lqcd{} curves.
Likewise, the sampled form factor fit includes no derivatives at all.
Since the fits with anchor points produces less favorable \pvalue{s}
 than the nominal fit, and because the nominal fit also has larger uncertainties
 in the $Q^{2}\lesssim0.1~\GeVsq$ range where the tension is seen,
 this slight difference is ignored.

\begin{figure}[htb!]
\includegraphics[width=0.48\textwidth]{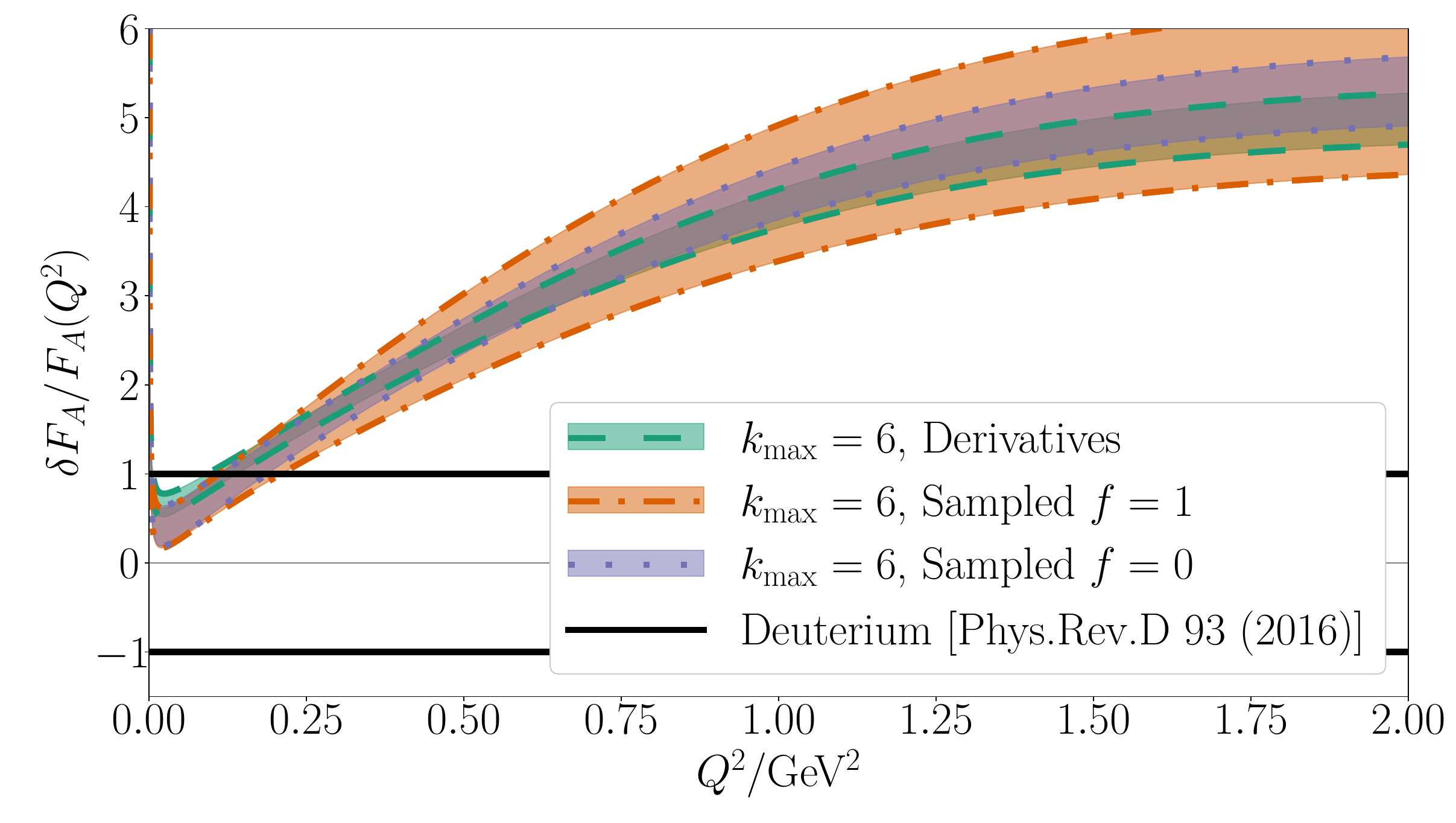}
\caption{
 Plot of the axial form factor fits from this work
 normalized by the \zexp{} deuterium result from \rfr{Meyer:2016oeg}.
 The teal dashed region corresponds to the nominal fit for the derivatives
 fit method.
 The orange dot-dashed region and the blue-violet dotted region
 correspond to sampled form factor fits,
 either with the assumption of unknown correlations being
 fully-correlated ($f=1$) or uncorrelated ($f=0$), respectively.
 For reference, the normalized range $0\pm1$ of the \zexp{}
 result is shown as black lines.
\label{fig:method_nominal}
}
\end{figure}

\subsubsection{Removing Sum Rules}
\label{sec:fitting_sumrules}

\fgr{fig:method_sumrules} shows the effects of dropping the
 sum rule constraints regulating the large $Q^{2}$ behavior from the fits.
The sampled form factor fits both take $f=0$ for uncorrelated results.
The fits all use 2 free parameters,
 giving $\kmax=6$ for fits with 4 sum rules
 or $\kmax=2$ for fits without sum rules.
The constraint fixing the value $F_{A}(Q^{2}=0) = g_{A}$
 is still imposed for all fits.

The fits both with and without sum rules agree well
 up to $Q^{2}\sim0.75~\GeVsq$.
At this point, the derivative fit with no sum rules starts to diverge
 from the full set of fits.
The sampled form factor fits with no sum rules start to
 diverge above $Q^{2}\sim1.0~\GeVsq$.
This is a reflection of the range of constraints on the form factor:
 for the derivatives fit, the largest $Q^{2}$ where the form factor is
 constrained comes from the \nme{} result with $-\tz=0.5~\GeVsq$;
 the sampled form factor fit has a $Q^{2}$ range extending up to $1.0~\GeVsq$.
Beyond these values, the form factor coefficients are not effectively constrained.
Without the sum rules to regulate the form factor,
 the form factor shape is free to exhibit divergent behavior.
However, in the range where fit constraints exist,
 the form factor should be described at least as well
 as without the sum rules provided \kmax{} is large enough
 to obtain a reasonable fit.

The sum rules do not introduce a large bias into
 the \zexp{} coefficients.
For the derivatives method, the fit coefficients with sum rules are
\begin{align}
 \big( a_{1}, a_{2} \big)
 = \Big(
 -1.701(42), 0.26(9)
 \Big) ,
\end{align}
 and without sum rules are
\begin{align}
 \big( a_{1}, a_{2} \big)
 = \Big(
 -1.726(77), -0.27(20)
 \Big) .
\end{align}
For the sampled form factors, the fit with sum rules gives
\begin{align}
 \big( a_{1}, a_{2} \big)
 = \Big(
 -1.630(63), 0.18(13)
 \Big) ,
\end{align}
 and without sum rules
\begin{align}
 \big( a_{1}, a_{2} \big)
 = \Big(
 -1.623(65), 0.02(21)
 \Big) .
\end{align}
This gives less than a $1\sigma$ shift in the value of $a_{1}$,
 but a more significant shift of $2.5\sigma$ for $a_{2}$ between
 the derivative method fits with and without sum rules.

\begin{figure}[htb!]
\includegraphics[width=0.48\textwidth]{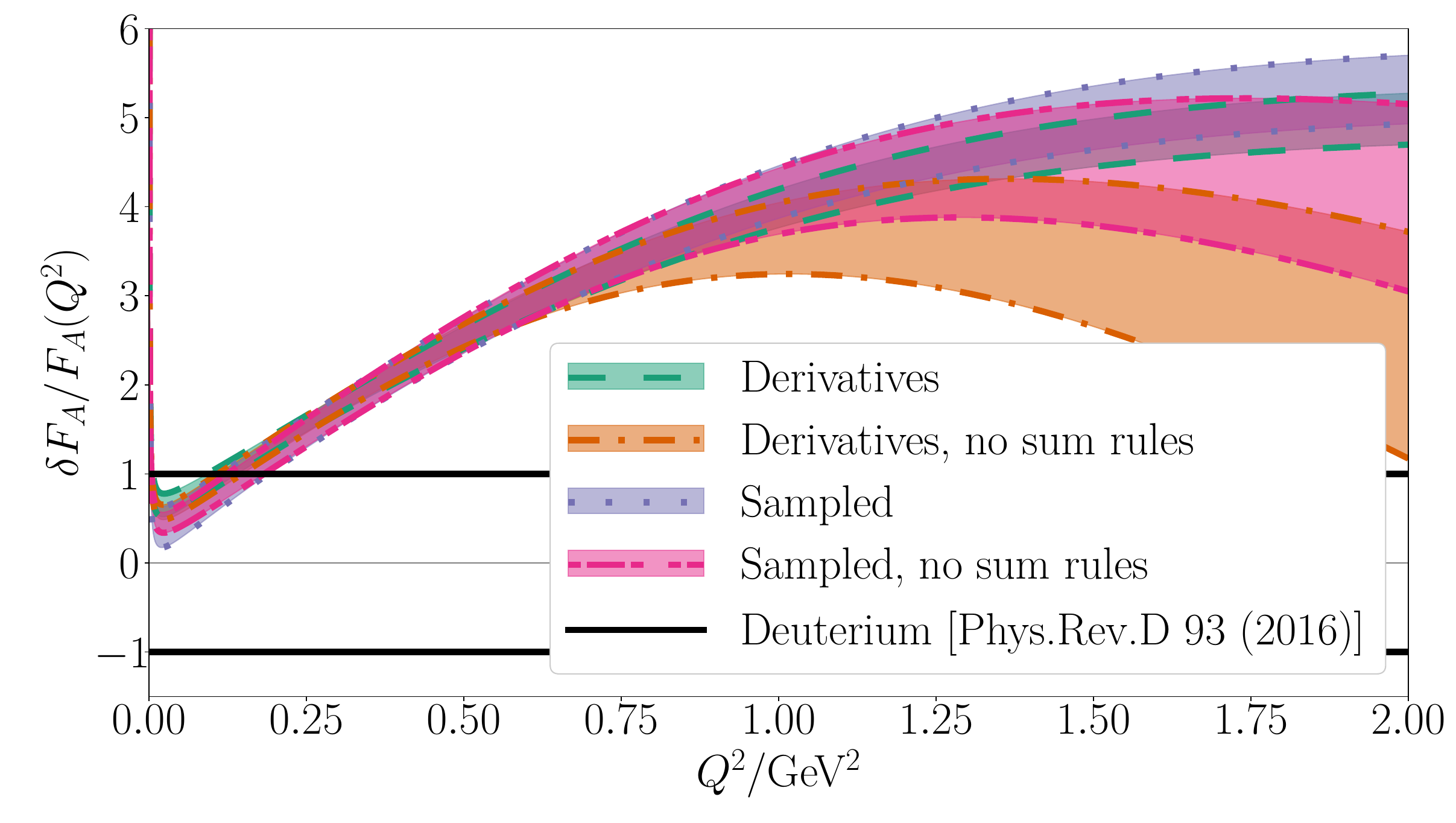}
\caption{
 Plot of the axial form factor fits from this work
 normalized by the \zexp{} deuterium result from \rfr{Meyer:2016oeg}.
 The teal dashed and orange dot-dashed regions correspond to fits with
 the derivatives method,
 while the blue-violet dotted and pink triple-dashed regions
 correspond to the sampled form factor fits.
 The teal dashed region and blue-violet dotted regions
 include the full set of four sum rules,
 and the orange dot-dashed and pink triple-dashed regions
 do not include any sum rule constraints.
 For reference, the normalized range $0\pm1$ of the \zexp{}
 result is shown as black lines.
\label{fig:method_sumrules}
}
\end{figure}

\subsubsection{\texorpdfstring{\tz}{Zero-Point} Variation}

The zero point in the definition of $z$, denoted by \tz,
 is another fixed parameter that should not make
 a significant difference in the fit.
These variations are explored in \fgr{fig:method_t0},
 where fits to the nominal $\tz=-0.50~\GeVsq$
 and a variation $\tz=-0.25~\GeVsq$ are plotted.

The value $\tz=-0.50~\GeVsq$ is selected to match
 onto a value optimized for the \minerva{} results
 in \rfr{Meyer:inprep}.
However, the range of $Q^{2}$ probed by \lqcd{} calculations
 only extends up to $Q^{2}\sim1~\GeVsq$.
The larger magnitude \tz{} means that $|z|\lesssim0.34$
 over the range $0\leq Q^{2} < 1~\GeVsq$,
 versus for $|z|\lesssim0.25$ for $\tz=-0.25~\GeVsq$.
This means that the form factor could potentially converge
 with fewer parameters if the smaller magnitude of $\tz$ was selected.

No substantial deviations are seen in the entire $Q^{2}$ range.
This is supported by the extracted \zexp{} coefficients as well.
The nominal fit with $\tz=-0.50~\GeVsq$ yields the values
\begin{align}
 \big( a_{1}, a_{2} \big)
 = \Big(
 -1.701(42), 0.26(9)
 \Big) .
\end{align}
To compare to this, the translated \zexp{} coefficients,
 defined by the prescription in \sct{sec:zexp_translated},
 for the modified value $\tz=-0.25~\GeVsq$ are
\begin{align}
 \big( a_{1}, a_{2} \big)
 = \Big(
 -1.771(36), 0.35(8)
 \Big) ,
\end{align}
 which agree to the level of $1.3\sigma$ for $a_{1}$ and $1\sigma$ for $a_{2}$.
For the sampled form factors, the fit with $\tz=-0.50~\GeVsq$ gives
\begin{align}
 \big( a_{1}, a_{2} \big)
 = \Big(
 -1.630(63), 0.18(13)
 \Big) ,
\end{align}
 and the fit for $\tz=-0.25~\GeVsq$ gives the translated values
\begin{align}
 \big( a_{1}, a_{2} \big)
 = \Big(
 -1.678(59), 0.22(12)
 \Big) ,
\end{align}
 which correspond again only to $1\sigma$ shifts.

\begin{figure}[htb!]
\includegraphics[width=0.48\textwidth]{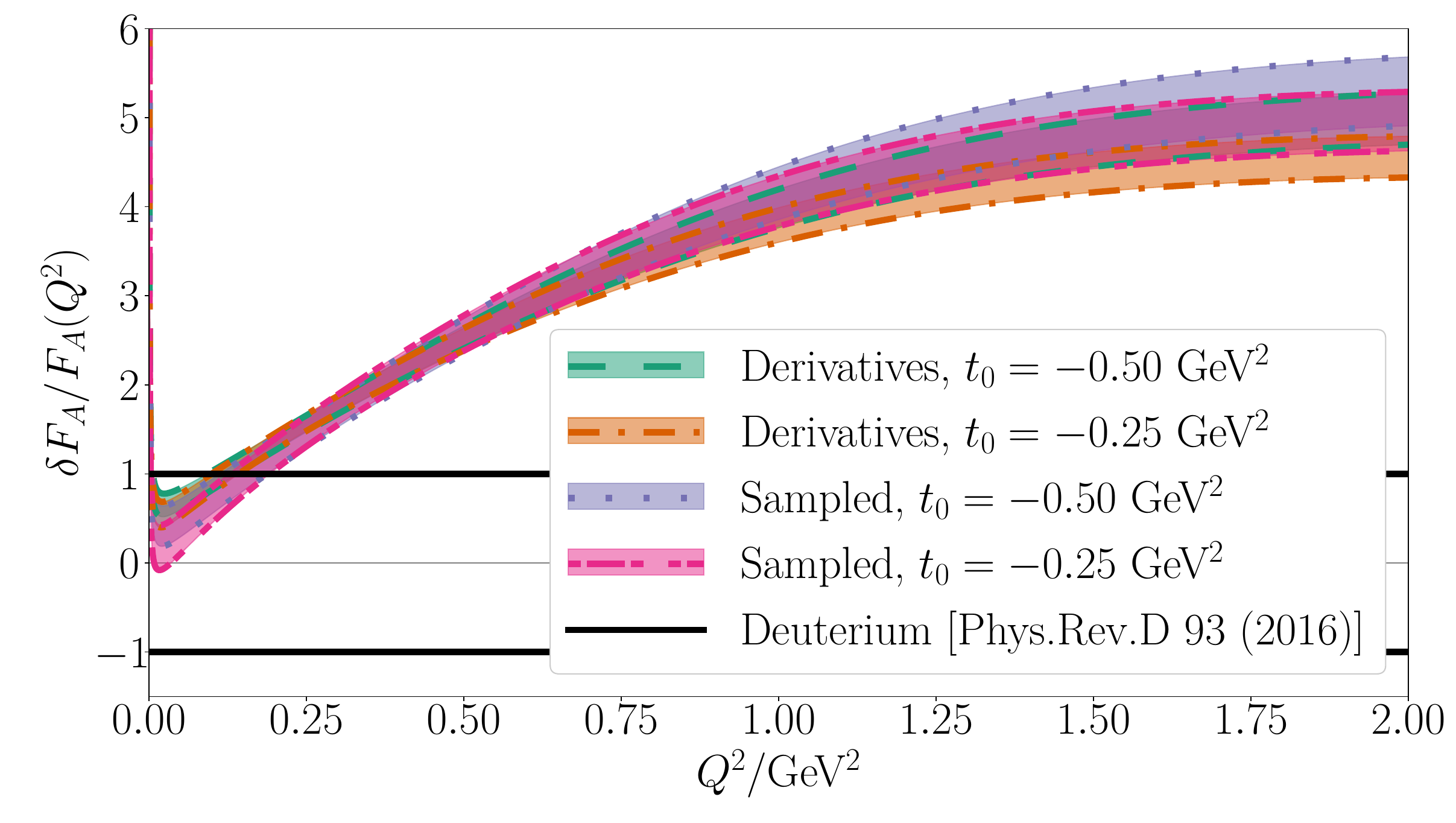}
\caption{
 Plot of the axial form factor fits from this work
 normalized by the \zexp{} deuterium result from \rfr{Meyer:2016oeg}.
 The teal dashed and blue-violet dotted regions show the fits to $\tz=-0.50~\GeVsq$
 with the derivatives and sampled form factor fit methods, respectively.
 The orange dot-dashed and pink triple-dashed regions correspond to
 fits with $\tz=-0.25~\GeVsq$ for the derivatives and sampled methods, respectively.
 For reference, the normalized range $0\pm1$ of the \zexp{}
 result is shown as black lines.
\label{fig:method_t0}
}
\end{figure}

\subsubsection{Fits to Dipole Form}

The \lqcd{} results are fit to a dipole form factor parameterization
 and compared to the nominal fits in \fgr{fig:method_dipole}.
The nominal fits for both the derivatives and sampled form factor methods
 are shown as the teal dashed and orange dot-dashed regions, respectively.
Superimposed is the fit to the dipole form factor,
 computed using the sampled form factor method.

The sampled form factor fit to the dipole parameterization
 yields an axial mass $M_{A}=1.236(38)~\GeV$.
The enhanced $M_{A}$ is typical of form factors with a slow $Q^{2}$ falloff,
 as is seen in the \lqcd{} results.
The dipole form factors depicted in \fgr{fig:method_dipole} exhibit
 deviations from the \zexp{} form factors,
 predicting a larger form factor for $Q^{2}\lesssim0.5~\GeVsq$
 and smaller for $Q^{2}\gtrsim1.0~\GeVsq$.
The reduced $\chi^{2}$ for the fit also does not perform
 as well as the \zexp{}, with a value of $\chi^{2}/\nu\approx1.24$
 versus the \zexp{} value $\chi^{2}/\nu\approx0.82$.

This figure demonstrates how the dipole parameterization fails
 to capture the form factor shape, even with an enhanced axial mass parameter.
The slow falloff of the form factor with $Q^{2}$
 is captured at the expense of a systematic bias in the form factor shape,
 first overpredicting the \zexp{} form factor at low $Q^{2}$
 then underpredicting relative to the \zexp{} at high $Q^{2}$.
The dipole parameterization does not have enough parametric freedom
 to change this trend, only to modify the value of $Q^{2}$
 where the crossover occurs.

\begin{figure}[htb!]
\includegraphics[width=0.48\textwidth]{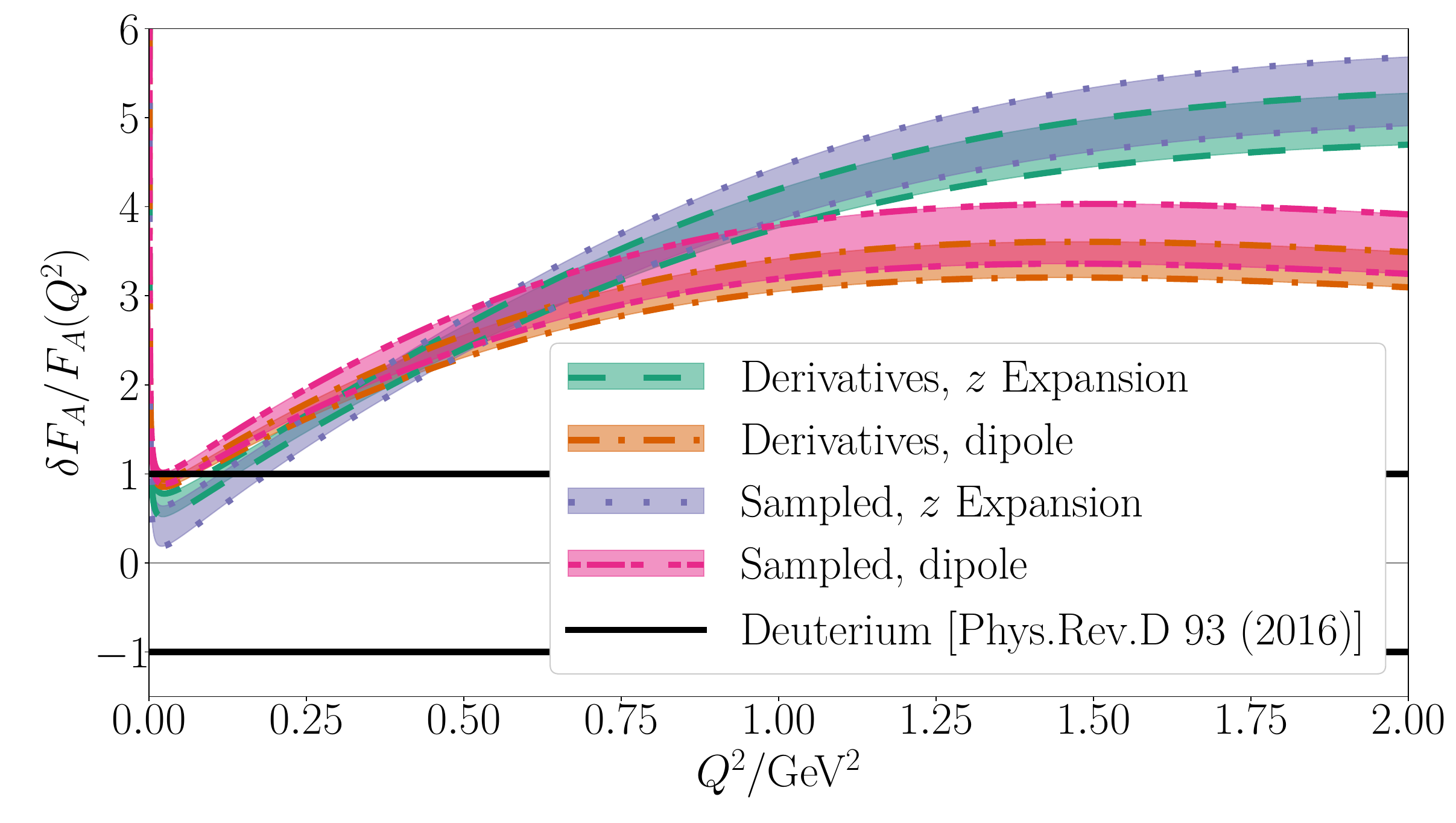}
\caption{
 Plot of the axial form factor fits from this work
 normalized by the \zexp{} deuterium result from \rfr{Meyer:2016oeg}.
 The teal dashed and blue-violet dotted regions show the fits to a \zexp{}
 parameterization with the derivatives and sampled form factor fit methods, respectively.
 The orange dot-dashed and pink triple-dashed regions correspond to
 dipole fits using the derivatives and sampled methods, respectively.
 For reference, the normalized range $0\pm1$ of the \zexp{}
 result is shown as black lines.
\label{fig:method_dipole}
}
\end{figure}

\subsection{Fit Parameterizations}

In this subsection, final fits of the form factors are given.
The values with two significant digits of uncertainty are provided
 for quick comparisons between fits.
In addition, the full set of parameter central values,
 including the parameters obtained from sum rule constraints,
 are provided along with the covariance matrix.
These values are provided for posterity
 so that the form factors can be used elsewhere.

\subsubsection{Fit to Derivatives}

The nominal fit to derivatives of the form factor is reproduced here.
The parameters fixed in the fits match those listed in \tbl{tab:fitparameters}.
The $\kmax=6$ fit is given by
\begin{align}
 \big( a_{1}, a_{2} \big)
 &=
 \big(
 -1.701(42), 0.263(90)
 \big)
 \label{eq:finalderiv6}
\end{align}
 with the covariance matrix
\begin{align}
 \left( \begin{array}{lll}
\phm0.00175929 &    -0.00294651 \\
   -0.00294651 & \phm0.00807158 \\
 \end{array} \right) .
\end{align}
With the sum rule constraints applied,
 the full set of coefficients are
\begin{align}
 &\big( a_{0}, \dots, a_{6} \big) \nonumber\\
 &=
 \begin{array}{clll}
 \big( & \phm0.72115656, &    -1.70104640, & \phm0.26324902, \\
       & \phm1.53433681, & \phm0.01061114, &    -1.49893610, \\
       & \phm0.67062898  & \big) . \\
 \end{array}
 \label{eq:finalderiv6_all}
\end{align}
The axial radius squared obtained from this parameterization is
\begin{align}
\rasq =  0.369(30)~\fmsq.
\end{align}
The unknown correlations between \lqcd{} results prevent
 the assignment of a definitive \pvalue.
Instead, covariance derating~\cite{Koch:2024tit}
 is used to estimate a $99\%$ confidence interval upper bound
 on the goodness-of-fit.
This limits $\p\lesssim\p^{\derated}\approx0.65$.
Assuming the unknown correlations are identically 0,
 the \pvalue{} can be computed directly as $\p^{\naive}\approx0.52$
 from $\chi^{2}\approx7.14$ and 8 degrees of freedom.

For a $\kmax=7$ fit with the same residuals as the $\kmax=6$ fit,
 the fit parameters are
\begin{align}
 \big( a_{1}, a_{2}, a_{3} \big)
 &=
 \big(
 -1.658(95), 0.431(347), 1.06(97)
 \big)
 \label{eq:finalderiv7}
\end{align}
 with the covariance matrix
\begin{align}
 \left( \begin{array}{lll}
 \phm0.00910412 & \phm0.02592048 &    -0.08986544 \\
 \phm0.02592048 & \phm0.12033231 &    -0.30490195 \\
    -0.08986544 &    -0.30490195 & \phm0.93874891 \\
 \end{array} \right) .
\end{align}
With the sum rule constraints applied,
 the full set of coefficients are
\begin{align}
 &\big( a_{0}, \dots, a_{7} \big) \nonumber\\
 &=
 \begin{array}{clll}
 \big( & \phm0.72240694, &    -1.65806627, & \phm0.43099261, \\
       & \phm1.06243035, &    -0.68256510, & \phm1.06363538, \\
       &    -1.59271086, & \phm0.65387696  & \big) . \\
 \end{array}
\end{align}
The axial radius squared obtained from this parameterization is
\begin{align}
\rasq =  0.341(62)~\fmsq.
\end{align}
For this fit, the \pvalue{s} are
 $\p^{\derated}\approx0.55$ and $\p^{\naive}\approx0.43$
 from $\chi^{2}\approx6.99$ and 7 degrees of freedom.

As an additional comparison point, the $\kmax=7$ fit
 with the second derivatives included is given by
\begin{align}
 \big( a_{1}, a_{2}, a_{3} \big)
 &=
 \big(
 -1.741(25), -0.048(49), 2.13(19)
 \big)
 \label{eq:finalderiv7anc}
\end{align}
 with the covariance matrix
\begin{align}
 \left( \begin{array}{lll}
 \phm0.00063056 & \phm0.00022664 &    -0.00454614 \\
 \phm0.00022664 & \phm0.00244803 &    -0.00422261 \\
    -0.00454614 &    -0.00422261 & \phm0.03561104 \\
 \end{array} \right) .
\end{align}
With the sum rule constraints applied,
 the full set of coefficients are
\begin{align}
 &\big( a_{0}, \dots, a_{7} \big) \nonumber\\
 &=
 \begin{array}{clll}
 \big( & \phm0.72415239, &    -1.74051938, &    -0.04845258, \\
       & \phm2.13430614, & \phm1.41235527, &    -5.65778627, \\
       & \phm4.15759466, &    -0.98165022  & \big) \\
 \end{array}
\end{align}
The axial radius squared obtained from this parameterization is
\begin{align}
\rasq =  0.423(16)~\fmsq.
\end{align}
For this fit, the \pvalue{s} are exactly equal with
 $\p^{\naive}=\p^{\derated}\approx0.32$
 from $\chi^{2}\approx13.73$ and 12 degrees of freedom.

\subsubsection{Fit to Sampled Form Factor}

The nominal fit to derivatives of the form factor is reproduced here.
The parameters fixed in the fits match those listed in \tbl{tab:fitparameters}.
The $\kmax=6$ fit is given by
\begin{align}
 \big( a_{1}, a_{2} \big)
 &=
 \big(
 -1.630(63), 0.18(13)
 \big)
 \label{eq:finalff6}
\end{align}
 with the covariance matrix
\begin{align}
 \left( \begin{array}{lll}
\phm0.00390684 &    -0.00480561 \\
   -0.00480561 & \phm0.01588552 \\
 \end{array} \right) .
\end{align}
With the sum rule constraints applied,
 the full set of coefficients are
\begin{align}
 &\big( a_{0}, \dots, a_{6} \big) \nonumber\\
 &=
 \begin{array}{clll}
 \big( & \phm0.72217245, &    -1.62716067, & \phm0.17896964, \\
       & \phm1.11227915, & \phm1.02836466, &    -2.30667668, \\
       & \phm0.89205145  & \big) \\
 \end{array}
\end{align}
The axial radius squared obtained from this parameterization is
\begin{align}
\rasq =  0.424(52)~\fmsq.
\end{align}

For a $\kmax=7$ fit with the same residuals as the $\kmax=6$ fit,
 the fit parameters are
\begin{align}
 \big( a_{1}, a_{2}, a_{3} \big)
 &=
 \big(
 -1.649(73), -0.06(27), 1.51(65)
 \big)
 \label{eq:finalff7}
\end{align}
 with the covariance matrix
\begin{align}
 \left( \begin{array}{lll}
 \phm0.00526364 & \phm0.00659750 &    -0.04317334 \\
 \phm0.00659750 & \phm0.07403803 &    -0.12297087 \\
    -0.04317334 &    -0.12297087 & \phm0.42623351 \\
 \end{array} \right) .
\end{align}
With the sum rule constraints applied,
 the full set of coefficients are
\begin{align}
 &\big( a_{0}, \dots, a_{7} \big) \nonumber\\
 &=
 \begin{array}{clll}
 \big( & \phm0.72415215, &    -1.64720287, &    -0.05571462, \\
       & \phm1.49700975, & \phm2.16783943, &    -5.42758266, \\
       & \phm3.45633330, &    -0.71483448  & \big) \\
 \end{array}
\end{align}
The axial radius squared obtained from this parameterization is
\begin{align}
\rasq =  0.476(79)~\fmsq.
\end{align}

\subsection{Comparison to \lqcd{} Parameterizations}

The fit result from \eqns~(\ref{eq:finalderiv6})--(\ref{eq:finalderiv6_all})
 is compared to the parameterizations it was fit to in \fgr{fig:compare_lqcd}.
The \lqcd{} calculations produce predictions of axial matrix elements
 at discrete, finite values of $Q^{2}$ in the range 0--1~\GeVsq,
 where the upper end of the range is limited only by finite computation resources.
These discrete values are fit across a range of lattice spacings,
 lattice volumes, and masses to produce the QCD predictions given here.
The \lqcd{} form factors are most precisely constrained at low-to-moderate $Q^{2}$,
 with the expectation that systematics will be largest at $Q^{2}$ close to 0.
The diverging form factor uncertainty for the \lqcd{} fits at $Q^{2}=0$
 is a consequence of the exact constraint $F_{A}(Q^{2}=0)=g_{A}^{\beta}$
 imposed on the deuterium result from \rfr{Meyer:2016oeg}.

The averaged form factor agrees well with the individual \lqcd{} results
 across the range of $Q^{2}$ shown in the figure.
None of the fit parameterizations include sum rules,
 so the agreement between the results is expected to degrade
 above the 1~\GeVsq{} cutoff.
Despite this expectation, the averaged form factor continues
 to follow along with all of the results except \rqcd{}
 above the 1~\GeVsq cutoff, at least in part due to the
 growing uncertainties at large $Q^{2}$.
In particular, the averaged form factor follows closely with
 the \nme{} result, which is the most precise of the available parameterizations.

The form factor uncertainty is improved significantly after the average.
Part of this constraint, especially at low $Q^{2}$ originates from the
 sum rule enforcing the condition $F_{A}(Q^{2}=0)=g_{A}^{\beta}$,
 which gives an infinitely precise prediction at $Q^{2}=0$.
However, the form factor is also about a factor of 2 more precise
 than the \lqcd{} results individually, which would be expected
 for an average of 5 uncorrelated results with similar uncertainties.
The precision of the form factor is about 1.2\% for $Q^{2}=0.5~\GeVsq$
 and 2.2\% at $Q^{2}=1.0~\GeVsq$.
This is near the limit of the precision that can be achieved
 without considering systematic corrections due to isospin-breaking
 effects, which are expected to enter at around the
 percent level~\cite{Cirigliano:2022hob}.

\begin{figure}[htb!]
\includegraphics[width=0.48\textwidth]{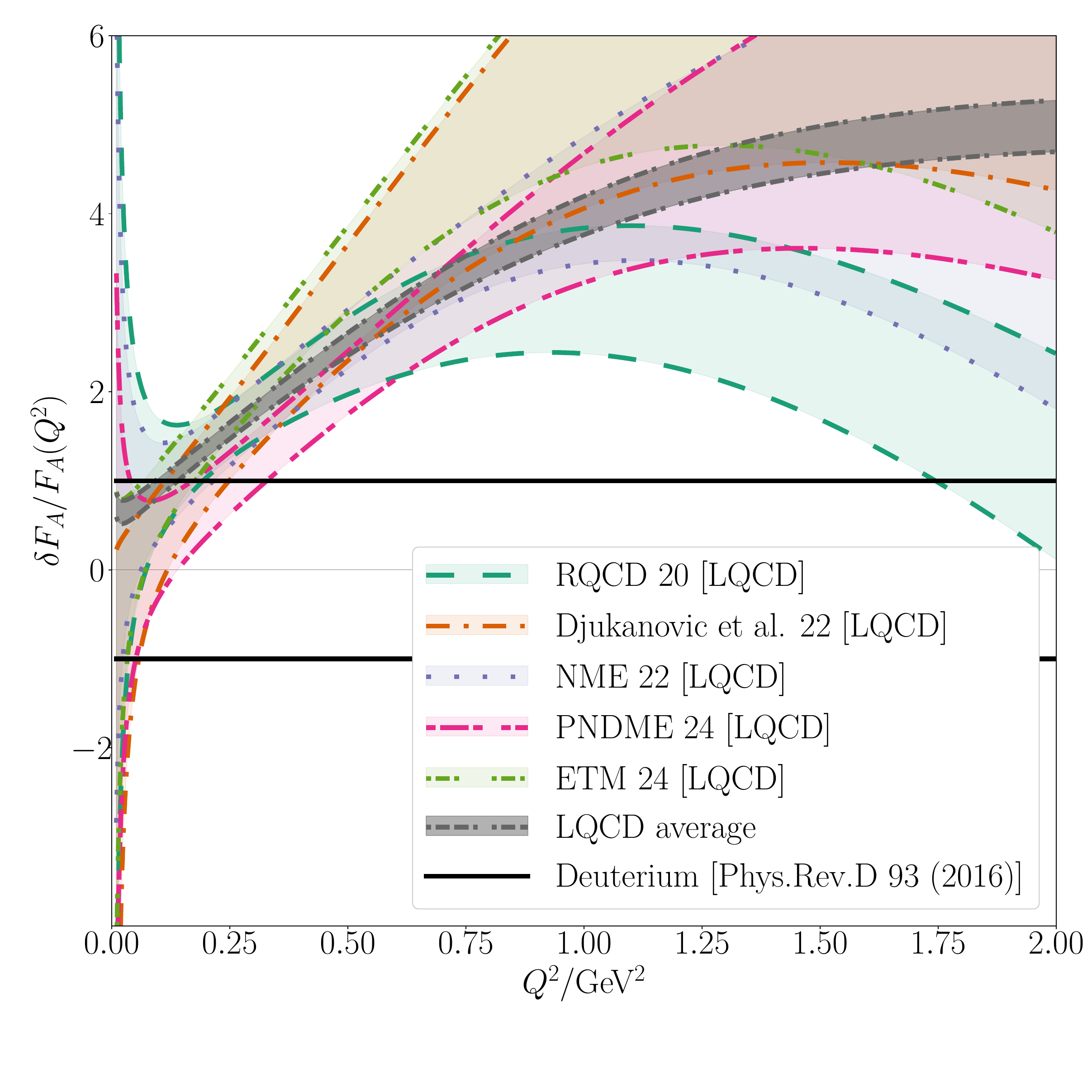}
\caption{
 Plot of the axial form factor results from the references
 in \tbl{tab:lqcd_comparison} normalized by the \zexp{}
 deuterium result from \rfr{Meyer:2016oeg}.
 The fit result from \eqns~(\ref{eq:finalderiv6})--(\ref{eq:finalderiv6_all})
 is shown as the gray region bounded by dot-dash-dash-dot lines.
 For reference, the normalized range $0\pm1$ of the \zexp{}
 result is shown as black lines.
\label{fig:compare_lqcd}
}
\end{figure}

\subsection{Comparison to Neutrino Scattering Data}

The comparison between the \lqcd{} average,
 neutrino-deuterium, and antineutrino-proton scattering data
 are shown in \fgr{fig:summary}.
The result from \rfr{Meyer:2016oeg} is shown for reference.
The fit to the deuterium bubble chamber data has implemented
 a different selection criterion for selecting the optimal
 amount of regularization for the \zexp{} parameters,
 appealing to an L-curve regularization scheme~\cite{Lcurve,Lcurvetext}
 versus arguments from unitarity.
This results in an approximate degeneracy between
 the floating normalizations on the deuterium cross section
 and the form factor shape, which leads to a large systematic uncertainty
 expressed as a dependence on the low-$Q^{2}$ cut.
The envelope of form factor shapes out to $1\sigma$ that are explored
 by different choices of $Q^{2}$ cut are given by the blue-violet region
 bounded by dotted lines.
With the increased uncertainty from the fit envelope due to systematic uncertainty,
 the deuterium does constrain the form factor shape strongly enough
 to compete with the \minerva{} or \lqcd{} results.

A fit to the \minerva{} antineutrino-proton dataset using the same set
 of \zexp{} parameters as the \lqcd{} is shown as the orange dot-dashed region.
Like the \lqcd{}, the \minerva{} fit exhibits a slower falloff with $Q^{2}$
 relative to the deuterium form factor.
There is still a slight tension, at the level of about $2\sigma$,
 between the \minerva{} and \lqcd{}.
A combined fit between the two sources yields a fit that is dominated
 by the \lqcd{}, lying almost entirely on top of the \lqcd{} curve.
This fit has been omitted from the figure.
A \pvalue{} test with a 1 \dof{} $\Delta \chi^{2}$ comparison
 demonstrates that the two form factor constraints are consistent
 despite the slight tension, yielding $\Delta \chi^{2} \approx 3.2$
 for a \pvalue{} of 0.08.
For a more comprehensive discussion of the connections to neutrino
 scattering data, the reader is referred to the sister paper, \rfr{Meyer:inprep}.

\begin{figure}[tbp]
\includegraphics[width=0.45\textwidth]{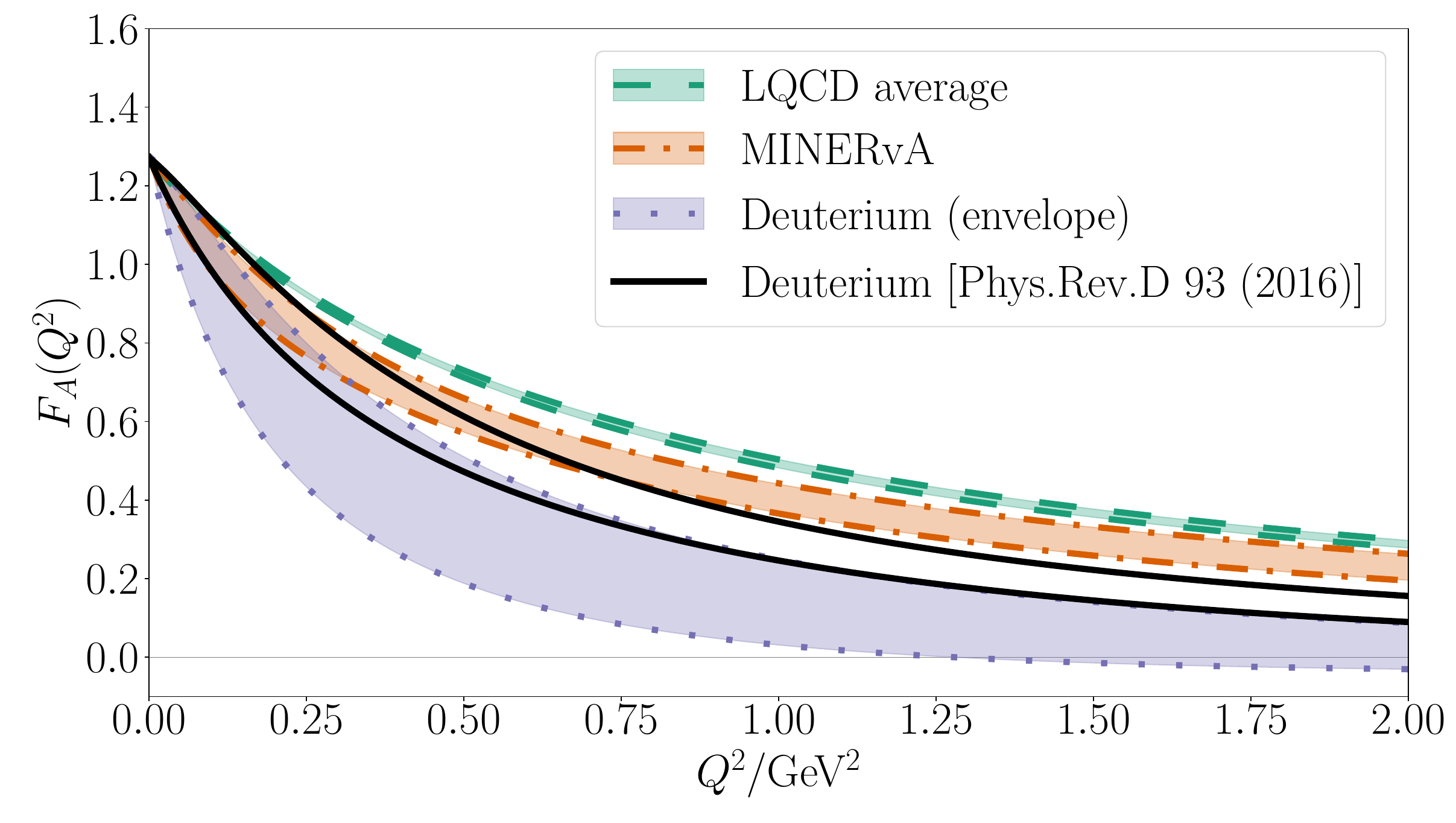} \\
\includegraphics[width=0.45\textwidth]{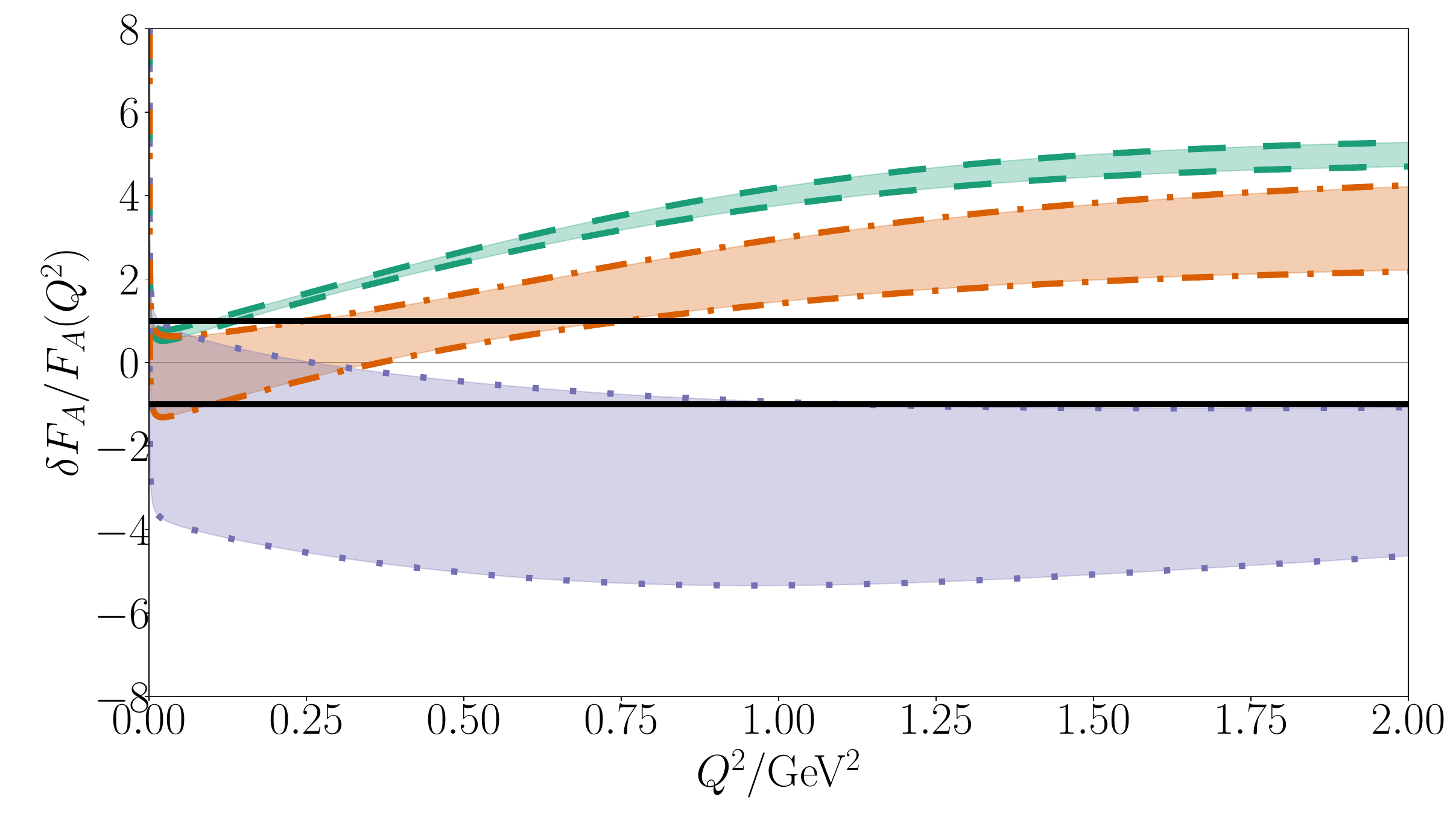}
\caption{
 The top panel shows the final choices for the
  axial form factor parameterizations taken from \rfr{Meyer:inprep}.
 The bottom panel shows those same parameterizations normalized to the
  deuterium result of \rfr{Meyer:2016oeg}.
 The \lqcd{} Average fit (green dashed region) is taken from \eqn{eq:finalderiv6}.
 The curve labeled \minerva{} (orange dot-dashed) is obtained from
  fitting the \zexp{} to antineutrino-proton scattering data
  in \rfrs~\cite{MINERvA:2022vmb,MINERvA:2023avz}.
 The curve labeled ``Deuterium (envelope)'' (blue-violet dotted) is taken
  from an envelope bounding two different kinematic cuts
  of fits to neutrino-deuterium bubble chamber scattering
  data~\cite{
 Mann:1973pr,
 Barish:1977qk,
 Barish:1978pj,
 Miller:1982qi,
 Baker:1981su,
 Kitagaki:1983px,
 Wachsmuth:1979th,
 Barlag:1984uga,
 Allasia:1990uy}
 For comparison, the result using neutrino-deuterium scattering data
  from \rfr{Meyer:2016oeg} is plotted as an unfilled region bounded by black solid lines.
 The bottom panel shows the same curves, but normalized by the
  result from \rfr{Meyer:2016oeg}.
 \label{fig:summary}
}
\end{figure}

\section{Discussion}
\label{sec:discussion}

Precise and accurate neutrino-nuclear cross sections are a necessity
 for next-generation neutrino oscillation experiments.
The nuclear models needed to predict these cross sections
 are constructed from neutrino-nucleon amplitudes.
These amplitudes are difficult to quantify from experiment alone.
As an example, charged-current neutrino-nucleon quasielastic scattering
 requires high-statistics measurements and detection of events
 with neutrons produced in the final state.
Given the relative importance of this low-energy interaction mechanism,
 it begs the question whether theory can be used to improve constraints
 on poorly-understood nucleon-level amplitudes.
Of particular interest is the nucleon axial form factor, which dominates
 the uncertainty budget for neutrino-nucleon quasielastic scattering.

\lqcd{} is a highly attractive candidate for providing theoretical inputs
 to neutrino-nuclear cross section predictions.
Calculations performed with \lqcd{} confer a number of advantages,
 such as rigorous uncertainty quantification
 and systematically improvable uncertainties.
Recent \lqcd{} calculations of the nucleon isovector axial form factor
 now pass key internal consistency checks, and also provide a consensus
 on the behavior of the form factor with respect to the momentum transfer dependence.
This behavior is in conflict with fits to neutrino-deuterium scattering results
 but consistent with antineutrino-hydrogen scattering,
 suggesting that remnant nuclear effects may still be at play even in
 deuterium scattering measurements.

\lqcd{} results also provide precise determinations of the form factor,
 leading to the appealing possibility of averaging the results to obtain
 an averaged \lqcd{} isovector axial form factor more precise
 than the individual results.
This possibility is explored in this work.
Two strategies for averaging the results are employed to perform the averages:
 first, an averaging strategy that fits to derivatives of the form factor
 with respect to $Q^{2}$; and second, a strategy that stochastically
 samples the form factor values over a set of equidistant $Q^{2}$ evaluation points.
Both strategies are compared to each other and average fits are produced.

The fit to the derivatives of the form factor takes advantage of the properties
 of the \zexp{} parameterization.
Defined as a power series of a small expansion parameter $z$,
 the \zexp{} naturally suppresses the effects of higher-order powers of $z$.
This leads to a form factor that can be approximately described by a linear
 function in $z$ over some range of $Q^{2}$, which suggests that the form factor
 shape can be reasonably well described by its central value and derivatives
 at a single point.
The ideal expansion point for this comparison is the value of \tz,
 which determines the value of $Q^{2}$ at which $z=0$.
This strategy for averaging works remarkably well and gives a result consistent
 with the typical strategy of fitting to the stochastically-sampled form factors
 in the average, as seen in \fgr{fig:method_nominal}.
This strategy can be defined in completely analytically
 with a fixed number of residuals, adding to the appeal
 and lending more direct assessments of goodness-of-fit
 and compatibility with experimental datasets.
The latter possibility is explored in a sister paper,
 \rfr{Meyer:inprep}.

The method of fitting to the derivatives has some drawbacks that must be considered.
Fits to the derivatives are sensitive to the $Q^{2}$ evaluation points that are used,
 particularly in constructing a fit with an invertible covariance matrix.
The form factor cannot be taken to arbitrarily large \kmax{},
 limited instead by the number of derivatives that are included and the spread
 of $Q^{2}$ evaluation points that are used.
If the evaluations are carried out only at the natural expansion point,
 i.e. evaluating the derivatives at $Q^{2}=-\tz$,
 then this strategy is also sensitive to the variety of \tz{}
 values provided by the included results.
Consistency is expected within the range of $Q^{2}$ explored by the \tz{} values,
 but beyond that range the form factor can drift away from the desired central value.
This is most apparent in \fgr{fig:method_sumrules},
 where the form factor drift in $Q^{2}$ can be plainly seen
 when the sum rules are not there to regulate the large $Q^{2}$ behavior.
This can be somewhat alleviated by including anchor points or sum rules,
 but in this work the former case also gives poor goodness-of-fit
 as reported in \tbl{tab:lqcd_comparison}.

The derivatives method also permits more sophisticated ways to estimate
 the uncertainty inflation needed to account for unknown sources of correlation.
The \lqcd{} results can be subject to unknown correlations over their statistical
 uncertainties due to reuse of gauge configurations.
In this work, the method of covariance derating introduced in \sct{sec:covariance_derating} 
 was used in the derivatives fitting method to assess the needed uncertainty inflation.
The derating method reports only a modest increase in uncertainty
 when allowing for covariance fluctuations with a 99\% confidence band,
 yielding a similar shape and precision as the sampled form factor
 with the assumption of no correlation between results
 as seen in \fgr{fig:method_nominal}.
This is a believable conclusion given that the results do not share
 their statistical samples, only (subsets of) the gauge configurations on which
 those statistical samples are generated.
With covariance derating, more guidance is provided on how to assign unknown correlations,
 allowing for a smaller uncertainty inflation when such an inflation is not needed.

The dipole parameterization shape visually disagrees with the results obtained
 from the \zexp{} fits, as seen in \fgr{fig:method_dipole}.
Although the slower falloff with $Q^{2}$ does result in an enhanced $M_{A}$
 (giving $M_{A}\approx1.21(2)~\GeV$ in \tbl{tab:results_fit_derivatives}),
 the dipole form factor is still larger than the \zexp{} below $Q^{2}\lesssim0.25~\GeVsq$
 and then smaller than the \zexp{} above $Q^{2}\gtrsim0.75~\GeVsq$.
This falls below several of the \lqcd{} results for $Q^{2}\gtrsim1.00~\GeVsq$.
Even still, the derivatives fit in \tbl{tab:lqcd_comparison} yields
 a \pvalue{} that is acceptable according to the criteria used
 to judge the \zexp{} parameterizations.

With the averaging strategy put forth in this work,
 the \lqcd{} results can be fit together with the \minerva{}
 antineutrino-hydrogen scattering data.
These fits produce a result that is nearly degenerate with
 the \lqcd{} result in isolation.
The fits from \lqcd{} and \minerva{} are compatible with each other,
 as determined from a \pvalue obtained from a $\Delta \chi^{2}$ test
 with 1 \dof{}.
The \lqcd{} and \minerva{}, both independently and together,
 disagree with the results from neutrino-deuterium scattering,
 suggesting an uncontrolled systematic in the deuterium scattering data.
These comparisons are explored in more detail in the sister paper
 to this work, \rfr{Meyer:inprep}.
With these considerations in mind,
 the recommendation is to opt for the \lqcd{} results
 relying on the confidence gained from comparisons between
 \lqcd{} and \minerva{} results.

\section{Acknowledgments}

I would like to thank
 Sara Collins,
 Peter Denton,
 Rajan Gupta,
 Andreas Kronfeld,
 Kevin McFarland,
 Sungwoo Park,
 Sasha Tomalak,
 Andr\'e Walker-Loud,
 Callum Wilkinson,
 and
 Clarence Wret,
 for useful discussions.
Special thanks goes out to Lukas Koch for providing guidance for many of the details
 in \sct{sec:covariance_derating} and additional functionality
 in the \nustattools~\cite{nustattools} package for the
 practical implementation of this work.

This work was performed under the auspices of
 the U.S. Department of Energy by Lawrence Livermore National Laboratory
 under Contract DE-AC52-07NA27344
 and the Neutrino Theory Network Program under Grant DE-AC02-07CHI11359
 and U.S. Department of Energy Award DE-SC0020250.

\appendix

\bibliographystyle{apsrev4-2}
\bibliography{main}

\end{document}